\documentclass[nonacm,sigplan]{acmart}

\usepackage[]{hyperref}

\usepackage{graphicx}
\usepackage{amsmath} 
\usepackage{float} 
\usepackage{enumitem}
\usepackage{multirow}
\usepackage{subcaption}
\usepackage{algorithm}
\usepackage{algorithmic}
\usepackage{overpic}

\usepackage{tikz}
\usepackage{circledsteps}

\newcommand{\note}[2]{{\color{#1}{[#2]}}}

\definecolor{darkgreen}{RGB}{0,128,0}
\newcommand{\neeraja}[1]{{\note{red}{NJY: #1}}}
\newcommand{\prasoon}[1]{{\note{blue}{P: #1}}}

\newcommand{\system}{\texttt{iServe}}

\newcommand{\submodel}{fingerprint}
\newcommand{\Submodel}{Fingerprint}

\begin{document}

\title{{\system}: An Intent-based  Serving System for LLMs}

\author[1,\dag]{Dimitrios Liakopoulos}
\authornote{Both authors contributed equally to this research.}
\affiliation{%
  \institution{The University of Texas at Austin}
  \city{Austin}
  \state{Texas}
  \country{USA}
}
\email{dimliak@utexas.edu}

\author{Tianrui Hu}
\authornotemark[1]
\affiliation{%
  \institution{The University of Texas at Austin}
  \city{Austin}
  \state{Texas}
  \country{USA}
}
\email{tianrui@utexas.edu}

\author{Prasoon Sinha}
\affiliation{%
  \institution{The University of Texas at Austin}
  \city{Austin}
  \state{Texas}
  \country{USA}
}
\email{prasoon.sinha@utexas.edu}

\author{Neeraja J. Yadwadkar}
\affiliation{%
  \institution{The University of Texas at Austin}
  \city{Austin}
  \state{Texas}
  \country{USA}
}
\email{neeraja@austin.utexas.edu}

\begin{abstract}

Large Language Models (LLMs) are becoming ubiquitous across industries, where applications demand they fulfill diverse user intents.
However, developers currently face the challenge of manually exploring numerous deployment configurations --- combinations of parallelism and compression techniques that impact resource usage, latency, cost, and accuracy --- to meet these intents. 
Assessing the impact of these configurations on user metrics requires extensive, costly profiling for each model. 
Existing approaches avoid this expense by using fixed, static configurations, but this often leads to sub-optimal performance and higher costs.
Moreover, none of these solutions dynamically adapt to changing user intents to balance latency and cost, effectively.

We present iServe, an automated, intent-based system for distributed LLM inference. 
Instead of manually selecting deployment configurations, developers simply specify their intent—such as minimizing latency, reducing cost, or meeting specific targets for either. 
iServe introduces \emph{fingerprints}, lightweight representations of LLMs, to efficiently estimate how different configurations impact latency and memory usage. 
Based on these insights and GPU availability, iServe dynamically selects the optimal configuration to align with the user’s intent.
For various LLMs and query arrival rates, 
{\system} best meets user intents compared to state-of-the-art systems by reducing latency by 77.62\% and SLO violations by 7.09$\times$ while improving GPU throughput by 4.72$\times$. Moreover, {\system}'s fingerprint-based profiling reduces profiling cost by 6.05$\times$ (GPU-hours) compared to baselines.

\end{abstract}


\maketitle 
\pagestyle{plain} 

\vspace{-2mm}
\section{Introduction}
\label{sec:introduction}

Large Language Models (LLMs) are set to become ubiquitous across industries where applications demand that they meet varied user intents. 
For instance, real-time chatbots and voice assistants prioritize low latency to deliver instant responses while education platforms and automated FAQs focus on cost-efficiency.
Meanwhile, mobile apps and IoT devices require minimal GPU usage.
To satisfy these diverse intents, experts introduced various (a) compression techniques that reduce the size of the models (e.g., quantization~\cite{dettmers2022llmint8, frantar2023gptq, xiao2023smoothquant, dettmers2023qlora, Yu20248bit, lin2023awq}, pruning~\cite{ma2023llmpruner, sun2023wanda, frantar2023sparsegpt}) and (b) parallelism techniques that distribute the models across multiple GPUs (e.g.,  data~\cite{shallue2019measuring}, pipeline~\cite{huang2019gpipe}, and tensor parallelism~\cite{Wang_2022}).
However, despite existing work, 
developers must manually search through hundreds of deployment configurations --- combinations of these techniques that impact resource footprints, latencies, costs, and accuracies --- to satisfy the user intents.

\begin{figure}[t]
     \centering
     \includegraphics[width=\linewidth]{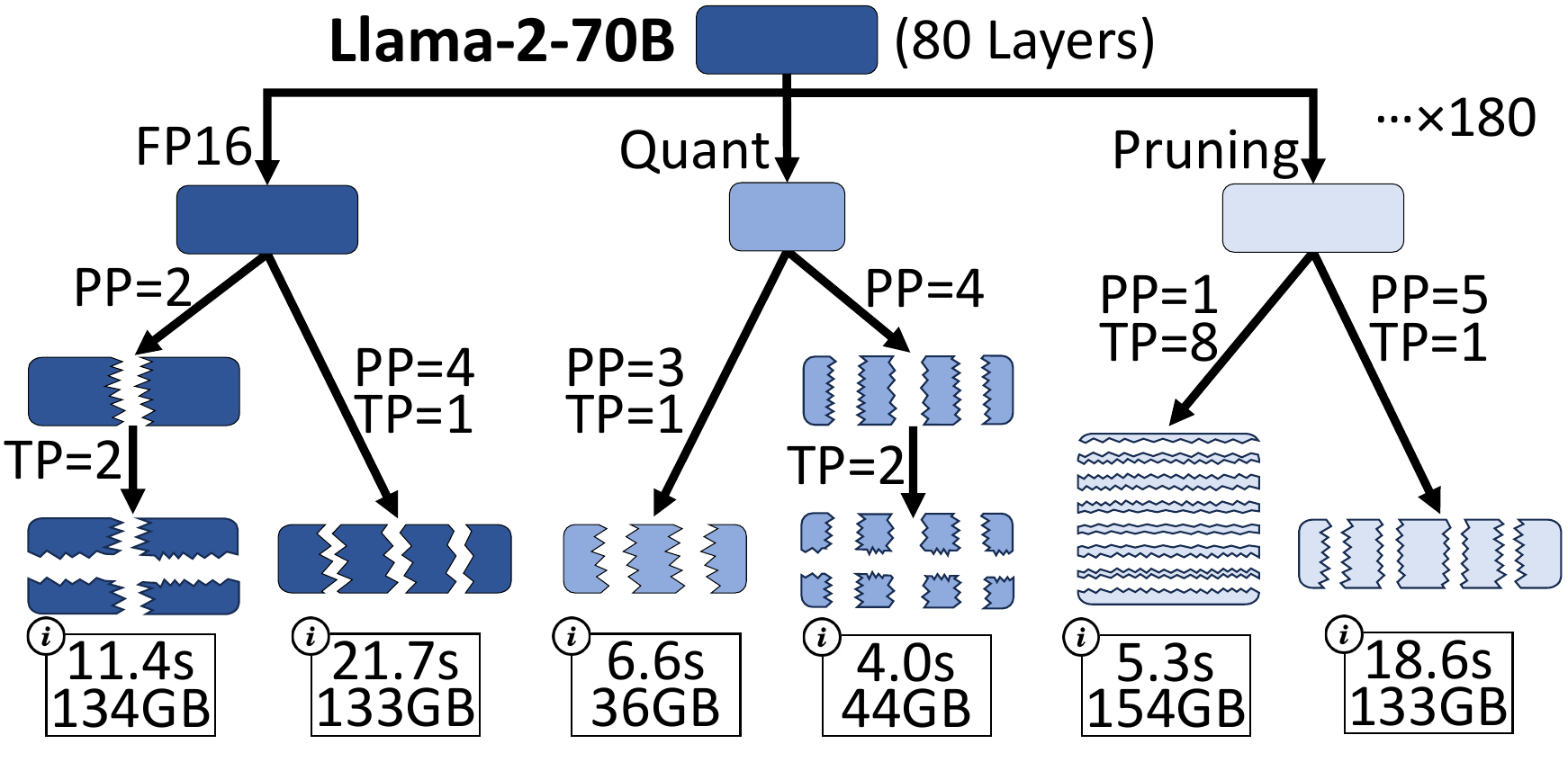}
     \vspace{-9mm}
     \caption{Each LLM has hundreds of deployment configurations with trade-offs in latency and memory consumption. The best configuration is specific to the user intent. See \S~\ref{sec:introduction}.
     }
     \vspace{-4mm}
     \label{fig:llama-70b-search-space}
 \end{figure}

The hundreds of resulting deployment configurations for each LLM vary across many dimensions. 
For example, Llama2-70B can be deployed in 180 ways on an 8-GPU cluster (Figure~\ref{fig:llama-70b-search-space}): 15 parallelization options (4 tensor, 8 pipeline) combined with 4 weight quantization methods and 2 KV-cache quantizations, or 4 pruning strategies. 
The inference latency for generating 100 tokens ranges from 
$2.58$ to $22.02$ seconds ($8.53\times$), memory consumption from $34.52$ to $153.59$ GB ($4.45\times$), and GPU time from $7.33$ to $174.37$ GPU-seconds ($23.79\times$). Finally, the hourly user cost for serving queries, based on the Azure LLM serving trace~\cite{patel2024splitwise}, ranges from $\$9.75$ to $\$268.40$ ($27.53\times$).
Thus, \emph{the choice of a deployment configuration for an LLM is key in meeting varied user intents.}

However, understanding the impact of deployment configurations on user metrics requires extensive and expensive profiling of each LLM. 
For example, it took $\sim$ 6 hours and 8 GPUs 
to profile a subset (80) of 
Llama2-70B's configurations.
Prior work circumvents this cost either by fixing the model parallelism techniques across different models~\cite{oh2024exegpt, zheng2022alpa, duan2024mopar} or leaving it to users to specify the deployment configuration~\cite{aminabadi2022deepspeed, tensorrt-llm, lin2023superscaler, kwon2023vllm, Athlur2022Varuna, CTranslate2}.
Both result in sub-optimal performance and cost for users, as (1) the best deployment configuration is LLM-specific, intent-specific, and depends on the availability of resources in the GPU cluster (\S~\ref{sec:characterization}), and (2) 
understanding the impact of configuration choices requires expertise in LLM architectures and hardware accelerators, which many users lack.

While some systems automate configuration selection by focusing only on parallelism techniques~\cite{li2023alpaserve, Galvatron2022, Agrawal2024, miao2023spotserve, Lin2024QUART, oh2024exegpt}, they (1) assume knowledge of future query arrival patterns~\cite{li2023alpaserve, Agrawal2024}, (2) require code changes to the LLM~\cite{Agrawal2024}, and/or (3) rely on expensive profiling~\cite{li2023alpaserve}.
Lastly, \emph{none adapt to varying user intents to trade off latency vs cost.}

\noindent\textbf{Our work.} 
We build {\system}, an automated intent-based system for distributed inference serving using LLMs. 
{\system} exposes an intuitive interface where application developers express the user intent: minimize latency or cost, or meet a specific target for latency or cost. 
{\system} introduces LLM \emph{fingerprints} to efficiently profile LLMs to estimate the impact of each deployment configuration on user metrics. 
{\system} then navigates the large deployment configuration search space for each LLM, on behalf of the developers, to select a deployment configuration to best meet the specified intent.
{\system} introduces a load-aware LLM-to-GPU placement algorithm to map the selected deployment configuration to the available GPUs in the serving cluster. 

Our key insight is that commonly used LLMs---decoder-based LLMs including GPT-J~\cite{gptj}, Llama-2~\cite{touvron2023llama}, Falcon~\cite{almazrouei2023falcon}, Mixtral MoE~\cite{jiang2024mixtral}---consist of multiple hidden layers with the \emph{same structural components}.
Using this insight, we introduce LLM \emph{fingerprints}, a reconstructed model that encompasses the same unique components (embedding layers, hidden layers, linear layers, softmax activations) as its LLM counterpart, but with at most two hidden layers (Figure~\ref{fig:fingerprint}).
Fingerprints capture the essence of the full LLM while being extremely lightweight: the memory footprint of Llama-2-70B's fingerprint is only 2.40\% of the entire LLM's.
Hence, instead of profiling the LLM, iServe profiles its fingerprint on a few configurations to estimate latency and memory for all deployment options.
Our fingerprint-based profiling significantly reduces profiling costs (time and resources), enabling {\system} to harvest spare resources in the GPU cluster for profiling instead of dedicated offline GPUs.

At deployment, {\system} uses the LLM's estimated data 
to select a configuration that fits available GPU memory and best meets user intent.
{\system} then applies a load-aware LLM-to-GPU placement algorithm to balance packing LLMs on fewer GPUs for resource efficiency versus spreading them out to minimize contention.

We build {\system} on top of TensorRT-LLM, a popular LLM inference serving platform widely used in industry~\cite{aws2024, alibaba2024, cerebrium2024, trtAmazon, trtAmex, trtZoox} and prior work~\cite{li2023speedodysseydeployablequantization, li2022easyefficienttransformer, wu2024fastdistributedinferenceserving, kwon2023vllm, 280922}.
Across a range of LLMs and query arrival rates from the Azure LLM inference serving traces~\cite{patel2024splitwise}, {\system} best meets user intent: {\system} reduces latency by 77.62\%, cost by 86.70\%, and SLO violations by 7.09$\times$ while improving GPU throughput by 4.72$\times$ on average compared to state-of-the-art inference serving systems~\cite{li2023alpaserve} and frameworks~\cite{sylvain2022accelerate,tensorrt-llm}. With its novel fingerprinting-based profiling, {\system} reduces profiling costs by 6.05$\times$  (GPU-hours), 
compared to AlpaServe~\cite{li2023alpaserve}.
We open-source {\system} at (link omitted for anonymity).

\vspace{-2mm}
\section{Background}
\label{sec:background}

\begin{figure}[t]
    \centering
    \includegraphics[width=.8\linewidth]{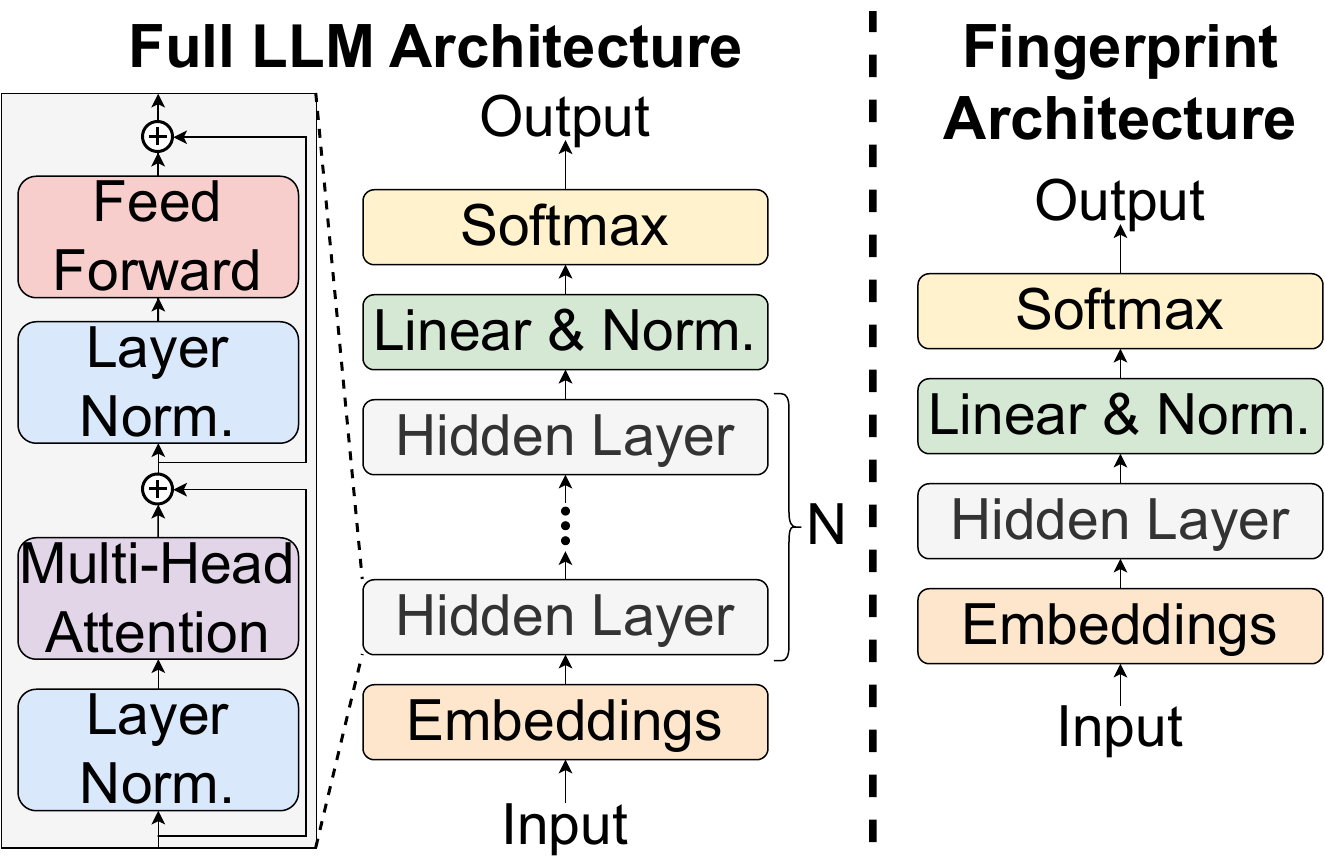}
    \vspace{-4mm}
    \caption{Architectures of a decoder-only LLM (left) and its {\submodel} (right). The {\submodel} has the same components as the full LLM, but only 1-2 hidden layers. See \S~\ref{sec:background}.}
    \vspace{-3mm}
    \label{fig:fingerprint}
\end{figure}

\noindent\textbf{Model architecture of LLMs.} 
LLMs are based on transformers, originally using encoder-decoder architectures (e.g., Bart~\cite{lewis2019bart}, Pegasus~\cite{zhang2020pegasus}, T5~\cite{raffel2023exploring}, UL2~\cite{tay2023ul2}) where the encoder creates fixed-sized vectors for the decoder to generate outputs.
However, GPT-2~\cite{radford2018improving} introduced a more flexible decoder-only architecture.
It showed that encoders are rigid, non-current structures that fix output lengths, whereas decoder-only architectures use recurrent structures that leverage previously generated tokens to generate outputs of arbitrary length.
Most modern LLMs now follow this decoder-only structure, stacking hundreds of identical layers to extract features (Figure~\ref{fig:fingerprint}).
Despite their size, the number of unique architecture components in these LLMs is small.


\noindent\textbf{LLM deployment configuration parameters.}
To efficiently deploy modern LLMs, expert users leverage three techniques: parallelism, quantization, and pruning.
Parallelism deploys LLMs across multiple GPUs: (1) data parallelism (DP)~\cite{shallue2019measuring} replicates LLMs across GPUs, (2) tensor parallelism (TP)~\cite{Wang_2022} splits tensor operations across GPUs, and (3) pipeline parallelism (PP)~\cite{huang2019gpipe} distributes layers across GPUs.
Parallelism techniques can be combined (e.g., TP and PP), however their extent is constrained by the number of available GPUs.

Unlike parallelism, quantization and pruning reduce an LLM's memory footprint.
Quantization~\cite{dettmers2022llmint8, frantar2023gptq, xiao2023smoothquant, dettmers2023qlora, Yu20248bit, lin2023awq} lowers the bit-width of LLM weight, activations~\cite{wu2023understandingint4quantizationtransformer}, and even KV-caches~\cite{int8kvcache, dong2024qaq}.
Pruning removes the redundant weights and layers of an LLM or optimizes LLMs for sparse acceleration~\cite{sun2023wanda, nvidia2021Sparse, ma2023llmpruner, frantar2023sparsegpt}.
While both slightly affect output quality, they are widely used for LLM deployment.

\vspace{-1mm}
\section{Motivation \& Characterization}
\label{sec:characterization}
\vspace{-1mm}

We study the impact of the deployment configuration parameters on four user intents: minimize inference latency, GPU memory consumption, cost (memory$\times$latency), and GPU-hours.
We then describe the challenges in selecting the optimal deployment configuration for an LLM.
We use an 8-GPU cluster with NVIDIA RTX A6000 (48GB) GPUs and study six state-of-the-art LLMs: Falcon-7B, Falcon-40B, GPT-J-6B, Llama-2-7B, Llama-2-13B, and Llama-2-70B. 

\vspace{-2mm}
\subsection{Effect of Parallelism}
\label{sec:parallelism}
\begin{figure}[t]
    \centering
    \includegraphics[width=\linewidth]{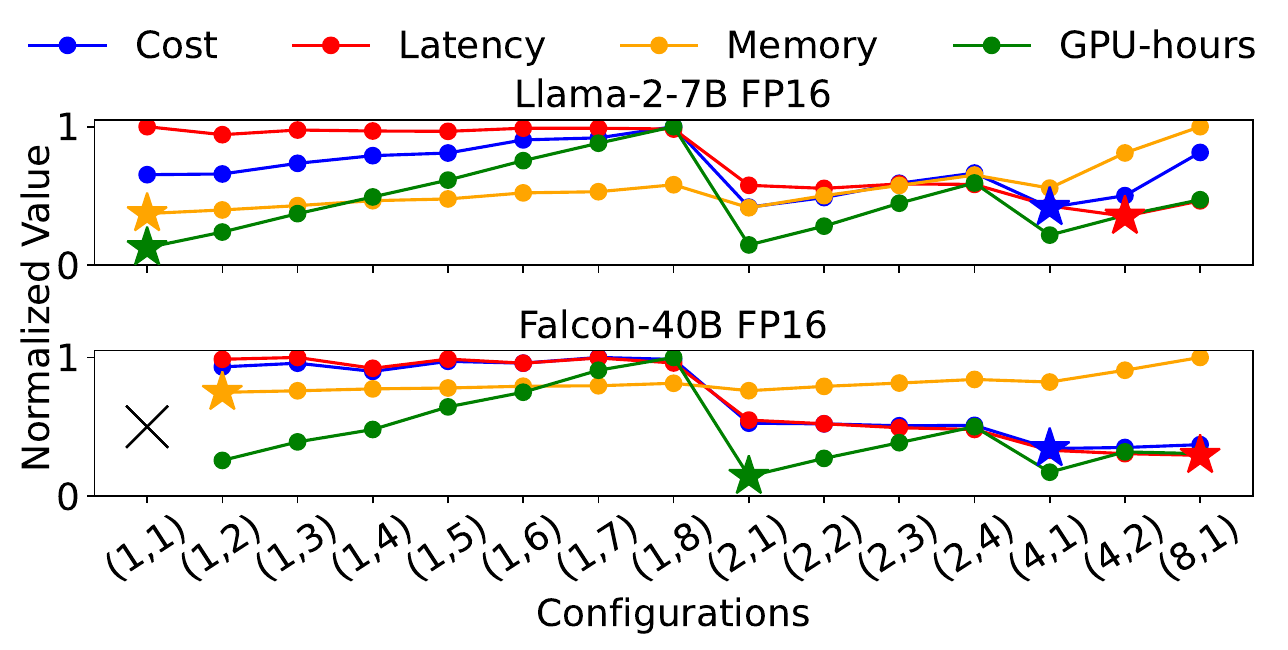}
    \vspace{-9mm}
    \caption{The best parallelism configuration (TP, PP) is user-intent and LLM-specific. See \S~\ref{sec:parallelism}.}
    \label{fig:tp-pp-curve}
\end{figure}

Figure~\ref{fig:tp-pp-curve} shows the cost, latency, total GPU memory, and GPU-hours as we adjust the parallelism (TP, PP) for Llama-2-7B and Falcon-40B. 


\noindent\textbf{Impact of PP alone.} 
Increasing PP from 1 to 8 ((1,1) to (1,8)) has minimal impact on the latency of a single query (0.02-0.86 sec difference), as the time to propagate intermediate data layer-by-layer between high-bandwidth devices is negligible ($<0.05\%$ the inference latency).
Total memory usage increases (1.01-1.56$\times$), but per-GPU memory decreases (1.48-5.14$\times$).
Thus, larger PP degrees may be useful in multiplexed environments where many GPUs (i.e., 8 for PP=8) have smaller fragments of free memory.

\noindent\textbf{Impact of TP alone.} 
Increasing TP from 1 to 8 
decreases latency while increasing the memory footprint and GPU-hours for both models: Falcon-40B's latency reduces by ($1.66$-$1.87\times$) while memory and GPU-hours increase by ($1.08$-$1.32\times$) and ($1.21$-$2.14\times$), respectively.
TP accelerates computation by parallelizing across devices, however, it also bloats the memory footprint and incurs an all-reduce latency overhead~\cite{shoeybi2020megatronlmtrainingmultibillionparameter}. 
For smaller LLMs like Llama-2-7B, the overheads of higher TP (four to eight) dramatically raise the cost by 1.95$\times$, while Falcon-40B only experiences a $1.08\times$ cost increase, making TP's impact on cost LLM-specific. 



\noindent\textbf{Analyzing combinations of TP/PP.}
Similar to our observations when analyzing TP and PP alone, the optimal combination of PP and TP is both LLM and intent-specific.
For example, while Llama-2-7B achieves the lowest latency with (4,2), Falcon-40B needs (8,1).
However, these configurations do not minimize other intents: for both LLMs, cost is minimized with (4,1), while memory is lowest with (1,2) for Falcon-40B and (1,1) for Llama-2-7B.


\noindent\textbf{\underline{Takeaway \#1.}} Larger PP degrees increase an LLM's total memory footprint but lower the per-GPU footprint. Increasing TP degrees reduces inference latency, however, excessive TP can deteriorate latency due to synchronization overheads. The best combination of TP/PP is LLM- and intent-specific.





\vspace{-2mm}
\subsection{Effect of Quantization}
\label{sec:quantization}
\begin{figure}[t]
    \includegraphics[width=\columnwidth]{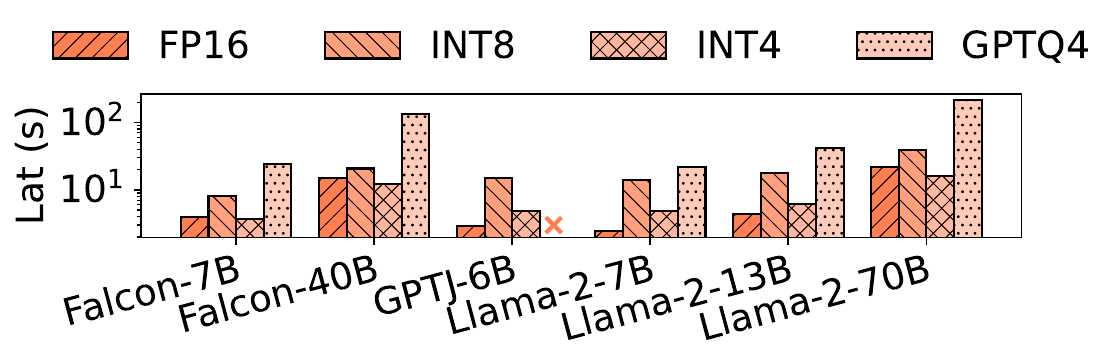}
    \vspace{-9mm}
    \caption{The impact of quantization on latency is LLM-specific. See \S~\ref{sec:quantization}.
    }
    \label{fig:quantization}
\end{figure}

The impact of quantization on memory is predictable (see supplementary material).
INT8 
cuts memory consumption by 50\%, while INT4 reduces it by 75\%.
GPTQ4 
incurs the least impact on output quality and reduces memory consumption by $\sim75$\%.


Unlike memory, the impact of quantization on latency is LLM-specific (Figure~\ref{fig:quantization}).
In PyTorch, smaller LLMs (Llama-2-7B, Llama-2-13B) perform best with no quantization, 
while INT4 reduces latency most for larger LLMs (Falcon-40B, Llama-2-70B).
GPTQ4 and INT8 incur higher latency.
Quantization reduces latency by streamlining matrix operations but increases it when weights are reconverted back to FP16 for activation.
For larger LLMs, INT4's reduced operations outweigh reconversion overheads.

\noindent\textbf{\underline{Takeaway \#2.}} 
Quantization predictably reduces an LLM's memory footprint, however, its impact on latency varies across LLMs.
There is no single optimal quantization technique: it is LLM- and intent-specific.

\vspace{-2mm}
\subsection{Effect of Pruning}
\label{sec:pruning}
\begin{figure}[t]
    \includegraphics[width=\columnwidth]{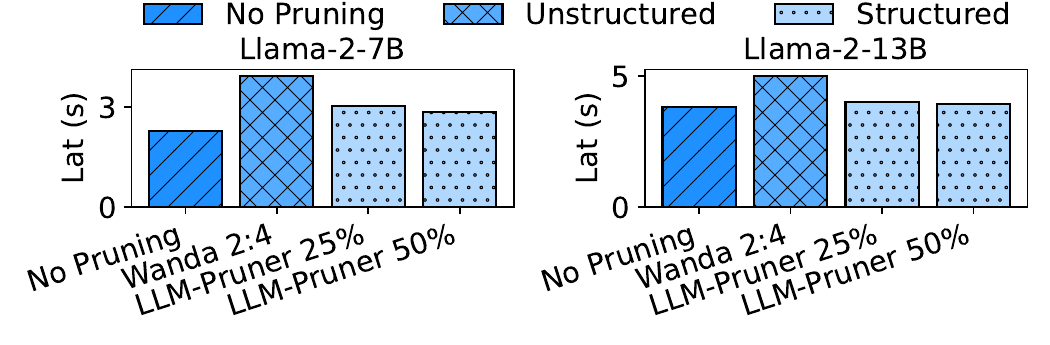}
    \vspace{-9mm}
    \caption{Pruning's effect on latency is LLM-specific. See \S~\ref{sec:pruning}.
    }
    \label{fig:pruning}
\end{figure}

We study the effects of (1) unstructured pruning~\cite{sun2023wanda} (zero out weights), and (2) structured pruning~\cite{ma2023llmpruner} (remove blocks of weights or layers) on memory and latency (Figure~\ref{fig:pruning}).
Both techniques reduce memory footprint.
Wanda 2:4 (unstructured) reduces memory by $\sim50\%$ by zeroing two of every four weights~\cite{nvidia2021Sparse}.
LLM-Pruner (structured) directly removes weights and layers: LLM-Pruner 25\% reduces the footprint by $\sim20\%$. However, their effect on latency is LLM-specific.
Unstructured pruning adds matrix conversion overheads~\cite{cai2023sparsity}, repeatedly converting sparse matrices to dense matrices and back on the critical path of computation, while the unbalanced weights in structured pruning poorly use the hardware~\cite{llm_pruner_issue33}.
Hence, both pruning techniques increase latency by 73.13\% and 33.04\% compared to no pruning for Llama-2-7B. 
However, for the larger Llama-2-13B, the latency increase is only 31.50\% for unstructured and 3.15\% for structured pruning, as the gains from reducing computational load with pruned matrices outweigh the overheads.

\noindent\textbf{\underline{Takeaway \#3.}} Pruning techniques present a trade-off between latency and memory consumption, however, the effect of pruning on inference latency is LLM-specific.

\vspace{-2mm}
\subsection{Challenges in Selecting a Configuration}

While we study each configuration parameter individually, their combinations complicate LLM deployment by expanding the search space and affecting latency and memory: 
for example, Llama-2-7B on an 8-GPU cluster has 158 unique configurations, with 82.14\% and 88.57\% variation in latency and memory, respectively.
While a single configuration optimizes each metric, limited GPU resources in dynamic clusters often prevent deploying the optimal one; hence, systems should understand each configuration's implications to select the best feasible deployment option.

Previous works make configuration decisions by either (a) using default configurations (e.g., Accelerate~\cite{sylvain2022accelerate}), or (b)
navigating the search space using expensive offline profiling~\cite{li2023alpaserve, Agrawal2024, miao2023spotserve}, or requiring oracle knowledge of the LLM load~\cite{li2023alpaserve, Agrawal2024}, and/or code changes to the developer's LLM~\cite{Agrawal2024}.
These design decisions are impractical in general-purpose systems with fixed GPU resources serving LLMs of non-expert users.
Profiling costs increase further with different LLM serving platforms (e.g., vLLM, TensorRT-LLM) as their underlying optimizations alter the latency/memory of each configuration.
Hence, new techniques are needed to efficiently navigate the search space and automatically select and deploy LLMs based on user intent.
\vspace{-2mm}
\section{{\system} Design}
\label{sec:design}

\begin{figure}[t]
    \centering
    \includegraphics[width=.9\linewidth]{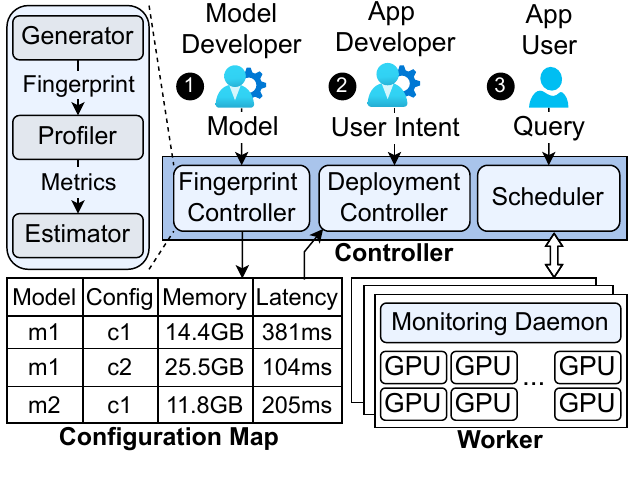}
    \vspace{-8mm}
    \caption{Architecture of {\system}. See \S~\ref{sec:design}.}
    \label{fig:system_design}
\end{figure}

We build {\system}, an automated intent-based LLM serving system. 
{\system} exposes an intuitive interface where developers express their intent: minimize latency or cost, or meet specific targets. 
{\system} navigates the deployment configuration space to select the best option and uses a load-aware placement algorithm to map the LLM to available GPUs.


\vspace{-2mm}
\subsection{Intent-Aware LLM Serving Interface}
\label{sec:interface}

LLMs are becoming ubiquitous across various domains where applications demand that they meet varied user intents. 
For instance, chatbots~\cite{openai2024chatgpt, dam2024completesurveyllmbasedai, Anthropic2023, microsoft_copilot_2024} and voice assistants~\cite{mahmood2023llmpoweredconversationalvoiceassistants, yang2024talk2care}, prioritize low latency to deliver instant responses while education platforms and automated FAQs
focus on cost-efficiency. Meanwhile, mobile apps and IoT devices require minimal GPU usage. 
However, to leverage LLMs, developers are forced to either (a) manually deploy them and reverse-engineer the best deployment configuration (TP, PP, quantization, pruning) that meets their intent, or (b) use rigid LLM inference serving systems for deployment that only support latency SLO specification~\cite{li2023alpaserve, sun2024llumnix, zhong2024distserve, Ahmed2024Proteus, Lin2024QUART, miao2023spotserve}.
While (a) burdens users with grasping the complex trade-offs of configuration parameters (\S~\ref{sec:characterization}), (b) burdens developers with choosing \emph{appropriate} latency SLOs and does not adapt to variable user intents.


{\system} is the first intent-based LLM inference serving platform that automatically deploys LLMs to meet variable user intents. 
{\system} unburdens developers by allowing them to simply specify their intent to minimize cost or latency without providing SLOs.
{\system} also supports developers who want to provide SLOs, either in the form of latency or cost.
Cost SLOs allow users to maximize their performance within budget constraints without reverse-engineering latency SLOs.
{\system} defaults to minimizing cost if no intent is specified.
As the cost formulation can vary from provider to provider, we evaluate {\system}'s efficacy with three cost formulations: (1) the traditional GPU memory $\times$ latency~\cite{netapp_fractional_gpu, truefoundry_fractional_gpus,mysticai_serverless_pricing}, (2) GPU memory consumed as cost, and (3) GPU-hours consumed as cost.
In this paper, we omit accuracy as a user intent, as the LLM community has yet to define a suitable metric for it~\cite{wang2024decodingtrust, hu2024perplexityreflectlargelanguage, jaiswal2024compressingllmstruthrarely, mo-etal-2024-trustworthy}.
We consider developing an LLM serving system that adapts to accuracy requirements as future work.

\vspace{-2mm}
\subsection{Workflow}
{\system} supports its intent-based interface with three main components (Figure~\ref{fig:system_design}).
The Fingerprint Controller profiles a registered LLM  with the LLM's \emph{fingerprints} to obtain metadata (i.e., latency and memory metrics) about the LLM across different configurations.
The Deployment Controller uses the LLM's profiled metadata to deploy the LLM with a configuration that can meet user intent (e.g., minimize cost). 
Finally, the Scheduler routes inference requests to the Workers in the GPU cluster hosting the LLM.
The three components are triggered based on the persona interacting with our system: LLM developers, application developers (e.g., ChatGPT), and application users (e.g., ChatGPT users). Figure~\ref{fig:system_design} details the workflows of {\system}'s three components.

\pgfkeys{/csteps/inner color=white}
\pgfkeys{/csteps/fill color=black}

\noindent\textbf{\Circled{1} {\system}'s model-registration workflow} is triggered 
when an \emph{LLM developer} submits an LLM to {\system}. 
The Fingerprint Controller takes charge: (a) the Generator produces the LLM's \emph{fingerprint}, a lightweight variant that encapsulates its unique model architecture in a smaller memory footprint. (b) 
The Profiler collects measurements from the fingerprint to estimate the latency and memory needs of the entire LLM. To do so, it deploys the generated fingerprint under a few deployment configurations and sends dummy inference requests to the fingerprint.
Being lightweight, the fingerprint allows {\system} to harvest available GPUs from the inference serving cluster for profiling, without negatively impacting the inference requests, rather than relying on dedicated GPUs. 
(c) 
The Estimator uses fingerprint-profiled data to estimate the latency and memory implications of the entire LLM for each deployment configuration.

\noindent\textbf{\Circled{2} {\system}'s model-deployment workflow} is triggered  
when an \emph{application developer} requests to deploy an LLM. They provide {\system} a user intent to optimize for. 
Using the configuration map with the estimated memory and latency for all configurations, {\system}'s Deployment Controller chooses the configuration that best meets the user intent and is feasible to deploy given resource availability.
It then uses a load-aware policy to select GPUs for deploying the LLM.



\noindent\textbf{\Circled{3} {\system}'s inference-serving workflow} is triggered  
when \emph{application users} submit inference requests to \system. 
The Scheduler dispatches them to worker nodes. 
We use a simple FCFS scheduler, as our focus in this work is on efficient profiling and deployment for LLMs to meet user intents. 
We design \system\ in a modular fashion to enable future work on developing advanced scheduling algorithms. 

\vspace{-2mm}
\section{Efficient Profiling using Fingerprints}
\label{sec:fingerprint_controller}
\begin{figure}[t]
    \centering
    \includegraphics[width=\linewidth]{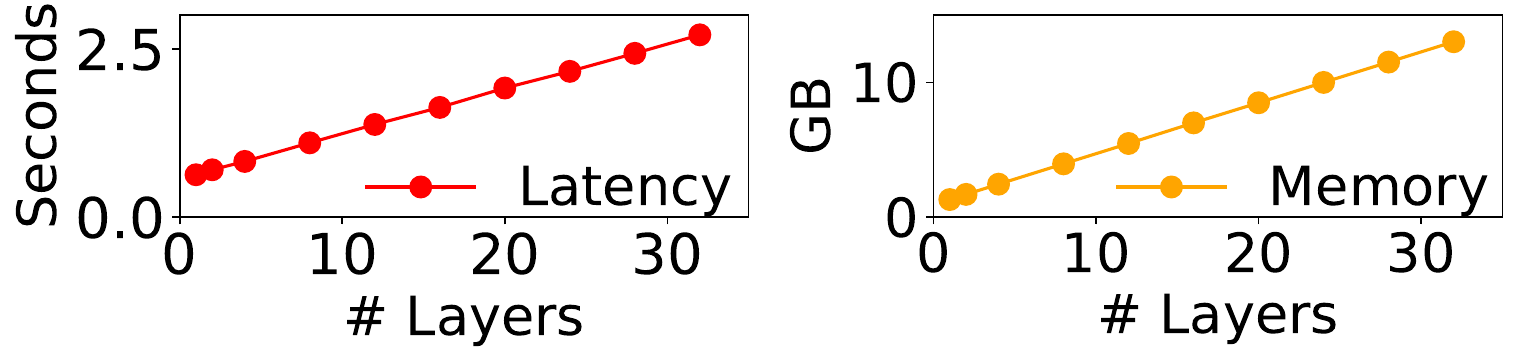}
    \vspace{-9mm}
    \caption{Latency and memory consumption are linearly proportional to the LLM's number of hidden layers. See \S~\ref{sec:fingerprint_controller}. 
    }
    \label{fig:fingerprint_scaling_new}
\end{figure}

Directly loading and profiling an LLM across all its deployment configurations is prohibitively expensive; it requires tens to hundreds of GBs of GPU memory and can take hours to profile.
The key components that bloat an LLM's memory footprint and inference latency are the identical hidden layers (\S~\ref{sec:background}).
We find a linear relationship between the number of hidden layers and memory footprint/inference latency (Figure~\ref{fig:fingerprint_scaling_new}): increasing an LLM's hidden layer count increases its memory footprint and inference latency by a predictable amount.
\emph{To reduce the profiling cost, we leverage the repetitive, predictable nature of LLM architectures.}

\vspace{-2mm}
\subsection{Generating LLM Fingerprints}
\label{sec:fingerprints_new}

Instead of profiling the full LLM, {\system} creates and profiles the LLM's \emph{fingerprint}, a lightweight variant capturing the key model architecture components: the embedding layer, hidden layers, normalization layer, and activation function (Figure~\ref{fig:fingerprint}).
Unlike the LLM, the fingerprint has minimal hidden layers, reducing memory and inference latency: Llama-2-70B's fingerprint is 41.59$\times$ smaller and 9.75$\times$ faster.


\vspace{-2mm}
\subsection{Estimating Memory} 
\label{sec:estimating_memory}
We now describe how we profile an LLM's fingerprint and extrapolate the memory requirements of the full LLM under every deployment configuration.

\begin{figure}[t]
    \centering
    \includegraphics[width=\linewidth]{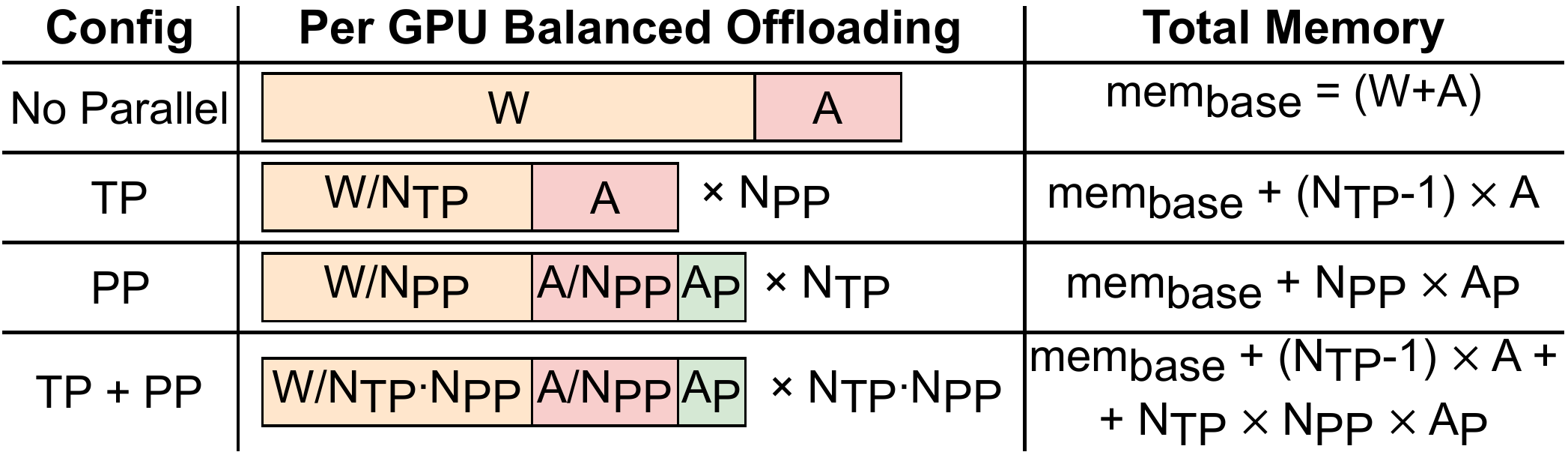}
    \vspace{-7mm}
    \caption{
    LLM parallelism memory overheads.
    Weights ($W$), activations ($A$), and pipeline activations ($A_p$) are replicated on $N_{TP}$ (TP degree) and/or $N_{PP}$ (PP degree) GPUs. See \S~\ref{sec:estimating_memory}.    
    }
    \label{fig:memory_overheads_new}
\end{figure}

\noindent\textbf{Breaking down an LLM's memory footprint.} 
Parallelism increases an LLM's memory footprint, as components are replicated across GPUs.
Figure~\ref{fig:memory_overheads_new} shows the memory footprint under different parallelism strategies~\cite{luo2023rtp}.
Three components define the footprint across parallelism schemes: weights ($W$), activations ($A$), and PP activations ($A_p$).
The memory footprint without parallelism ($mem_{base}$) is simply the weights and activations ($W+A$).
TP inflates this by $(N_{TP}-1) \times A$, where $N_{TP}$ is the TP degree, as each layer's activations are replicated on each GPU.
PP replicates the intermediate PP activations, inflating the base footprint by $N_{PP}\times A_{P}$, where $N_{PP}$ is the PP degree.
Finally, combining TP and PP incurs both overheads.
\emph{Knowing $W$, $A$, and $A_p$ allows us to estimate the LLM's memory footprint for any deployment configuration.}

\noindent\textbf{Candidate memory profiling techniques.}
We use our understanding of the memory footprint to devise three candidate methods (M1, M2, and M3) to estimate the memory footprint per configuration for an LLM.

\emph{M1. LLM under selected configurations}:
To reduce profiling costs compared to a brute force approach (observing the LLM under all configurations), we leverage our insight that knowing $W$, $A$, and $A_p$ allows us to estimate memory for any configuration.
To estimate these three values, we profile three configurations and solve a $3\times3$ system of equations (Figure~\ref{fig:memory_overheads_new}).
Because these values change with quantization (quantization can be applied to weights and/or activations), we profile three parallelism degrees per quantization: TP/PP=1, TP=1 PP=2, and TP=2 PP=1, which require the least resources.
This method reduces the profiling runs by only observing three configurations but still requires significant GPU memory.


\emph{M2. {\Submodel} under selected configurations}:
To reduce the profiling resource requirements, we profile the LLM's {\submodel}s (not the LLM) following M1's methodology.
However, solving the system of equations using the fingerprint's footprint yields $W$, $A$, and $A_p$ to be specific to the fingerprint, not the LLM.
Thus, we first linearly extrapolate (Figure~\ref{fig:fingerprint_scaling_new}) the LLM's footprint from the fingerprints under the three parallelism schemes.
Then, we solve the equations using the extrapolated LLM footprints to estimate $W$, $A$, and $A_p$.


\emph{M3. {\Submodel} under all configurations}:
While M2 reduces profiling cost, it relies heavily on extrapolation: it extrapolates the LLM’s footprint from the fingerprint and uses the extrapolated footprints to further extrapolate $W$, $A$, and $A_p$.
To reduce the potential inaccuracy from extrapolation, we consider one final technique.
We observe our fingerprint under every deployment configuration and linearly scale the LLM's footprint from the fingerprints.
This method trades off accuracy versus profiling cost compared to M2.

\vspace{-2mm}
\subsection{Estimating LLM Inference Latency}
\label{sec:latency-profiling_new}

Unlike memory, inference latency scales with LLM output length, making precise latency estimation during LLM registration difficult.
Instead, we capture key metadata (e.g., time to first token) per configuration to help trivially estimate latency during deployment time.
We outline our insights and two strategies for latency profiling.


\noindent\textbf{Breaking down an LLM's inference latency.}
LLMs generate outputs token by token using autoregressive decoding. 
After processing the input tokens for the first output, each subsequent token is generated.
The KV-cache avoids recomputing preceding tokens, ensuring consistent latency for each new token generation.
Thus, LLM inference latency is formally broken down into the time to first token (TTFT), time per output token (TPOT), and the output length~\cite{griggs2024melangecostefficientlarge, miao2023spotserve}:
\vspace{-3mm}
\begin{equation}
\label{eq:lat}
latency = TTFT + output\_length \times TPOT
\end{equation}
While output length is specific to the inference query, $TTFT$ and $TPOT$ are \emph{specific to the LLM's deployment configuration}.
Hence, knowing $TTFT$/$TPOT$ per configuration enables {\system} to estimate latency for any output length.
Our goal is to efficiently obtain these values for every configuration.


\noindent\textbf{Candidate latency profiling techniques.}
Similar to memory profiling, we use our understanding of LLM latency to devise two profiling strategies (L1 and L2) for estimation.



\emph{L1. LLM under selected configurations}:
Similar to memory, this method reduces profiling costs by observing the LLM under fewer configurations to obtain $TTFT/TPOT$ per configuration.
We find that we only need to observe configurations with PP=1 if we disable batching.
With batch size 1, PP overheads are minimal (<0.05\% for Llama-2-70B) and overheads from unbalanced layer assignments are eliminated~\cite{liu2022autopipe}.
This allows us to reuse $TTFT$ and $TPOT$ values from PP=1 for any configurations with PP$\geq2$.
While batching increases GPU utilization, these large LLMs already fully utilize GPUs at batch size 1.
Moreover, small batches mimic the behavior of LLM serving systems where latency-critical queries are processed sequentially.
Hence, similar to previous systems~\cite{li2023alpaserve}, we disable batching in this work.
For every configuration with PP=1, we observe the latency for two output lengths.
We then solve a $2\times2$ system of equations using Equation~\ref{eq:lat} to estimate the two unknowns $TTFT$ and $TPOT$.
Though this method reduces profiling runs, it still requires significant GPU memory.

\emph{L2. Fingerprints under selected configurations}:
While we can use L1's methodology to get the fingerprint's $TTFT$ and $TPOT$, these values are specific to the fingerprint, not the LLM with more hidden layers.
Therefore, we break down $TTFT$ and $TPOT$ into the latency through the LLM's building blocks: the hidden layers and the other layers (\S~\ref{sec:background}).
Formally, 
\begin{equation}
\label{eq:ttft}
TTFT = num\_layers \times TTFT_{layer} + TTFT_{other}
\end{equation}
\begin{equation}
\label{eq:tpot}
TPOT = num\_layers \times TPOT_{layer} + TPOT_{other}
\end{equation}
where \(num\_layers\) is the number of hidden layers in the LLM, \(TTFT_{layer}\) is the latency for one hidden layer to produce the first token, \(TPOT_{layer}\) is the latency for one hidden layer to produce an output token, and \(TTFT_{other}\) and \(TPOT_{other}\) are the latencies of the other basic building blocks for their respective token generation.
With knowledge of these four variables in Equations~\ref{eq:ttft} and ~\ref{eq:tpot}, we can compute the LLM's $TTFT$ and $TPOT$ with $num\_layers$ hidden layers per configuration.
Hence, in this method, we profile the LLM's fingerprints to efficiently obtain $TTFT_{layer}$, $TPOT_{layer}$, $TTFT_{other}$, and $TPOT_{other}$ per deployment configuration.

We observe two fingerprints per LLM, each with one and two hidden layers.
We send two requests with different output lengths to each fingerprint.
Then, for each fingerprint, we solve a $2\times2$ system of equations using Equation~\ref{eq:lat} to obtain its $TTFT$ and $TPOT$.
Using these values, we then solve a $4\times4$ system of equations using Equations~\ref{eq:ttft} and ~\ref{eq:tpot} to estimate $TTFT_{layer}$, $TPOT_{layer}$, $TTFT_{other}$, and $TPOT_{other}$ per configuration.
Similar to L2, we only observe configurations where PP$=1$ to reduce profiling runs.

\begin{figure}[t]
    \centering
    \includegraphics[width=\linewidth]{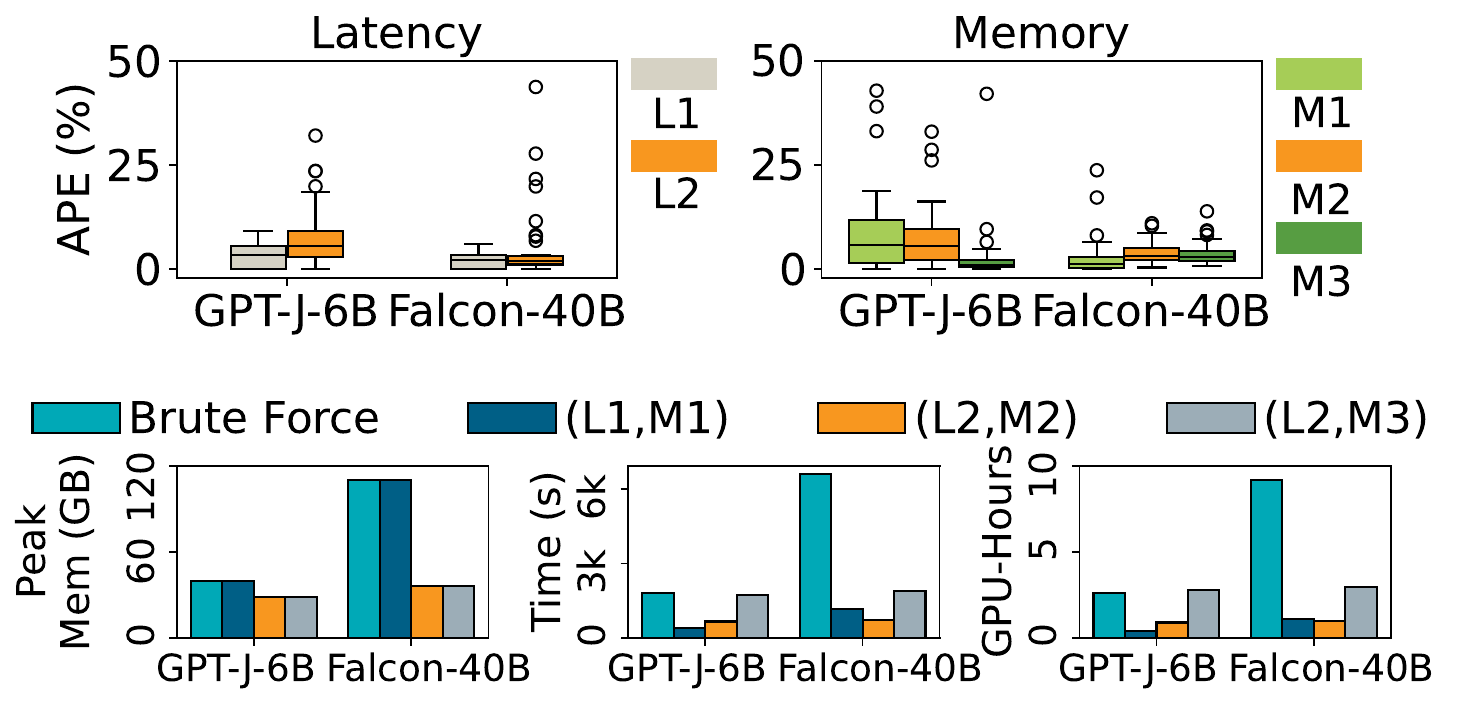}
    \vspace{-8mm}
    \caption{Design exploration of profiling methods. We show profiling accuracy (row 1) and cost (row 2) for each method. {\system} uses (L2,M2) to reduce profiling cost. See \S~\ref{sec:latency-profiling_new}.
    }
    \label{fig:mem_and_lst_prof}
\end{figure}

\noindent\textbf{Design exploration of profiling techniques.}
We evaluate the efficacy of the proposed profiling methods for two LLMs: GPT-J-6B and Falcon-40B.
Figure~\ref{fig:mem_and_lst_prof} reports profiling accuracy (in latency and memory) and cost (in profiling memory, time, and GPU-hours) for combinations of the memory and latency strategies.
We combine methods using the full LLM (M1 and L1) with those that use fingerprints (L2 with M2-M3).


Of the two fingerprint-based memory profiling methods (M2-M3), M3's accuracy is best: it observes every configuration's memory footprint, while M2 extrapolates from fewer observations.
M2's relative error is lower for larger models (e.g., Falcon-40B) than smaller models, as it consistently mispredicts by 1.2-1.9 GB on average.
L1 exhibits less variability in latency error by observing the full LLM, while L2's fingerprint profiling achieves median accuracy similar to L1, with $<2.10\%$ difference for both GPT-J-6B and Falcon-40B.


Profiling cost varies among combinations of the profiling strategies.
While L2 combined with M2-M3 exhibit the same peak GPU memory, (L2,M2) reduces profiling time by 61.6\% for GPT-J-6B and 60.68\% for Falcon-40B compared to (L2,M3) due to fewer observations.
We observe similar trends for GPU-hours.
While the profiling time with the full LLM (L1,M1) is similar to using fingerprints (L2,M2), the profiling memory usage is much higher when observing LLMs.
Ultimately, we deploy {\system} with the (L2,M2) profiling methods due to its lower profiling cost and relatively similar accuracy to other strategies.
However, our modular design allows operators to easily swap techniques.

\vspace{-2mm}
\section{Intent- and Load-Aware LLM Deployment}
\label{sec:deployer}

{\system}'s Deployment Controller makes two decisions: (1) which configuration to deploy an LLM with, and (2) which GPUs in the cluster to deploy the LLM on.
We devise a simple, greedy algorithm to make these decisions.

\noindent\textbf{Overview.} 
The Deployment Controller follows a simple, three-step algorithm.
(I) Using profiling-based estimates, it first ranks the LLM's deployment configurations that best meet the given user intent (e.g., minimize latency).
(II) It then determines the first feasible deployment configuration. 
(III) Finally, it maps the LLM to specific GPUs in the cluster. 
We describe these steps in detail below. 


\noindent\textbf{Determining a feasible deployment configuration.}
To determine the first feasible configuration to deploy an LLM with (step II), the Deployment Controller traverses the ranked list of configurations.
It first determines the memory requirement of a single layer of the LLM under the configuration:
\[mem_{block} = \frac{mem_{base}}{N_{TP} \times {num\_layers}}\]
where $N_{TP}$ is the TP degree, $mem_{base}$ is the memory footprint of the model from Figure \ref{fig:memory_overheads_new}, and \(num\_layers\) is the number of hidden layers in the LLM.
This basic memory block is configuration-specific, as it captures the memory scaling for a single layer due to the parallelism and quantization used. 

Then, the Controller selects the top $K$ GPUs with free memory ($K=TP \times PP$) to check if all of the LLM's memory blocks can be split across the $K$ GPUs in some partition (e.g., 2 on one GPU, 4 on all others) without exceeding each GPUs available memory.
We iteratively assign memory blocks from GPU 1 to GPU $K$, skipping any GPU exceeding its memory limit, until all blocks are assigned (and proceed to step III) or we move to the next configuration.


\noindent\textbf{Mapping LLMs to GPUs.}
Finally, the Controller maps the LLM to $K$ specific GPUs.
In the context of LLM serving, there is little work that explores different placement policies.
AlpaServe uses a brute-force algorithm to iterate over all placement possibilities~\cite{li2023alpaserve}, but profiles LLMs for $>1$ hour and needs a priori knowledge of request patterns, limiting its use for general-purpose LLM serving.
Other cloud systems (e.g., serverless~\cite{Kaffes2022Socc}) explore placement policies that account for fluctuating load but need fine-grained CPU core occupancy details, which is impractical for GPUs due to the high overheads of GPU toolchains like Nsight Compute ($>100\times$).


We evaluate the efficacy of three placement policies under fluctuating load. 
We define a GPU's load as:
\[load_{i} = \frac{\text{number of seconds GPU } i \text{ is busy}}{\text{window screening period}}\]
This definition is more expressive: it captures the amount of meaningful work the GPU completes, not just memory utilization or the number of loaded models.
We track GPU load every second over a 120-second window; this window size matches the rate at which load changes in production LLM clusters (every 2-3 minutes)~\cite{patel2024splitwise, stojkovic2024dynamollmdesigningllminference}.

\begin{figure}[t]
    \centering
    \begin{overpic}[width=.9\linewidth]{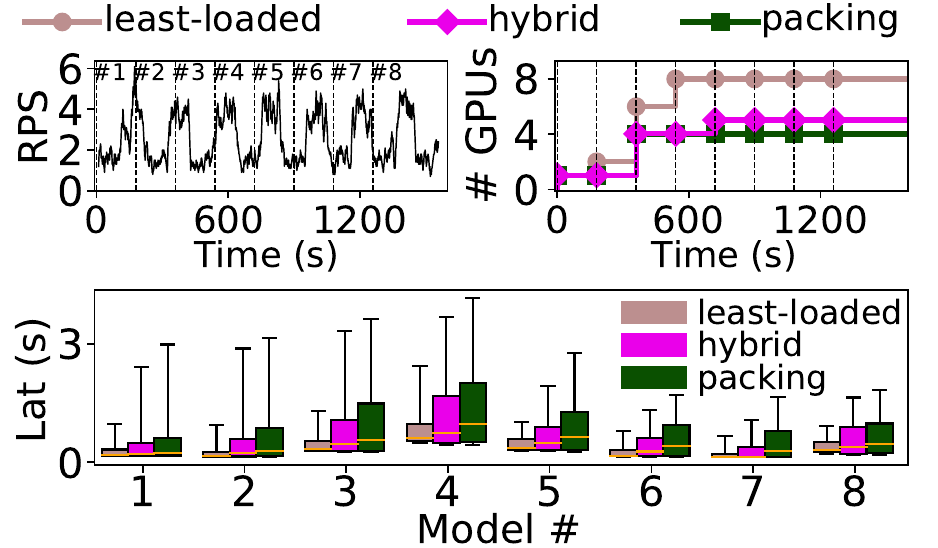}
    \end{overpic}
    \vspace{-4mm}
    \caption{Design exploration of placement policies. The hybrid policy balances packing LLMs onto fewer GPUs to reduce the \# of GPUs consumed and dispersing LLMs across GPUs to reduce latency. See \S~\ref{sec:deployer}.
    }
    \vspace{-2mm}
    \label{fig:deployment_policies}
\end{figure}

We consider three placement policies: least-loaded, packing, and hybrid.
Least-loaded deploys the LLM on the first $K$ least-loaded GPUs ($K=TP \times PP$) with enough free GPU memory, spreading LLMs across GPUs to reduce contention.
Packing deploys the LLM on the first $K$ GPUs with the highest load and enough free memory, prioritizing packing LLMs on fewer GPUs.
Hybrid is a flexible, load-aware placement policy.
We categorize the GPUs as packable ($load_{GPU}<\theta$) and non-packable ($load_{GPU}>\theta$), where $\theta$ is a load threshold. 
We first attempt to deploy the LLM on packable GPUs, following the packing policy.
If the per-GPU memory across the packable GPUs is insufficient, we deploy the LLM on non-packable GPUs following the least-loaded policy.

Figure~\ref{fig:deployment_policies} shows the inference latency (bottom) and number of GPUs used (top right) under the three placement policies.
Following the methodology of previous work~\cite{Kaffes2022Socc, li2023alpaserve}, we generate a 30-minute synthetic trace with fluctuating load using a Poisson process and deploy an LLM every 2-3 minutes to assess policy responses to load spikes and valleys (Figure~\ref{fig:deployment_policies} top left).
Least-loaded minimizes latency by dispersing LLMs: it uses all eight GPUs after deploying the fourth LLM.
Packing only uses four GPUs, however it inflates latency due to GPU contention.
Hybrid strikes a balance with five GPUs: during high load (e.g., M3 deployment), hybrid loads the model on idle GPUs rather than the previous, tightly packed one, but during low load (e.g., M4, M6-M8 deployment), it loads on packed GPUs.
This slightly increases median and p95 latency by 0.05-0.76$\times$ and 0.45-2.08$\times$, respectively, however, we justify the small latency trade-off for better GPU utilization. Thus, we use the hybrid policy but provide a knob to switch policies as needed.

\noindent\textbf{Loading the LLM under a deployment configuration.}
After selecting the configuration and GPUs, we assign memory blocks (layers) to each GPU; we keep adjacent layers close and distribute blocks evenly to reduce parallelism overheads~\cite{10.1145/3605573.3605627}. 
{\system}'s Deployment Controller then deploys the LLMs with these block assignments on the chosen GPUs.


\noindent\textbf{Scheduling.} 
We use a simple FCFS policy to schedule queries, as scheduling is orthogonal to {\system}; advanced schedulers (e.g., fair LLM scheduling~\cite{sheng2024fairness}) can easily be integrated into {\system}'s modular design.
Moreover, we exclude runtime preemption, swapping, and batching.
Preemption can improve SLO attainment~\cite{davis1993scheduling} and can easily be integrated into {\system}.
We avoid swapping as the latency overheads of offloading LLM weights to CPU DRAM or disk deteriorate inference latency and throughput~\cite{aminabadi2022deepspeed}; we evaluate against a state-of-the-art framework that employs swapping (\S~\ref{sec:evaluation_new}).
Finally, though batching is disabled, {\system}'s Profiler can easily predict LLM latency and memory needs for different batch sizes using simple modeling (e.g., linear regressors~\cite{li2023alpaserve, nexus}).

\vspace{-4mm}
\section{Implementation}
\label{sec:implementation}

We build {\system} in Python (3K LoC) on top of TensorRT-LLM~\cite{tensorrt-llm}, a widely-used LLM inference serving framework~\cite{trtAmazon, trtAmex, trtZoox}.  
We customize TensorRT-LLM to support unbalanced pipeline parallelism with specific model layer assignments. 
{\system}'s Controller is a persistent process that manages a pool of threads to handle requests made to {\system} depending on the persona interacting with {\system}.

\noindent\textbf{Fingerprint Controller.} 
The Fingerprint Controller runs as a CPU process.
Upon receiving an LLM registration request, it spawns a thread to generate and profile the LLM's fingerprint to capture latency and memory metrics per deployment configuration.
To create the fingerprint, we use PyTorch's API to read the LLM weight file containing LLM weights and architecture information (provided by the LLM developer in a commonly used safetensors format~\cite{safetensors}) and remove redundant layers.
Then, the thread deploys the fingerprint on GPUs to capture memory usage (via the monitoring daemon) and latency by issuing dummy requests.
We store the profiled data in a configuration map locally for efficient access by {\system}'s Deployment Controller.

\noindent\textbf{Deployment Controller.}
Upon an LLM deployment request, {\system}'s Controller creates a deployment thread that selects the deployment configuration and GPUs, and launches a TensorRT-LLM process on the worker for the LLM. The thread then stores the worker’s IP and port in an in-memory map, enabling the Scheduler to dispatch requests.

\noindent\textbf{Monitoring Daemon.}
{\system} launches a lightweight monitoring daemon per worker.
A single-threaded daemon on the worker’s CPU tracks GPU memory and load via the PyTorch CUDA API.
It sends updates to {\system}'s Controller every second via Linux TCP sockets~\cite{tcpmanpage}.
The Controller maintains a map of the available GPU memory and GPU load per server, which it leverages during model deployment.

\vspace{-3mm}
\section{Evaluation}
\label{sec:evaluation_new}
\vspace{-1mm}

\noindent\textbf{Workload.}
We evaluate {\system} using two Azure LLM inference traces~\cite{patel2024splitwise} that provide real-world query arrival patterns: (1) the \emph{conversation} trace with a steady, high query arrival pattern, and (2) the \emph{code} trace with bursty queries.
Following prior work~\cite{clockwork2020Gujarati, li2023alpaserve, Ahmed2024Proteus, miao2023spotserve}, we scale the traces to different query arrival rates while preserving the original pattern (e.g., burstiness).
We compress the trace by discretizing it into intervals, aggregating request counts from adjacent intervals, and scaling by a rate factor (0.1-0.4) to generate low to high load where GPU utilization is between 30-100\%.
The traces only include a timestamp and input/output size per query, but obfuscate the query input and intended LLM; we randomly assign each query to a model with a random input (max 1K tokens) from the ShareGPT dataset~\cite{sharegpt}.
We use six state-of-the-art LLMs (Llama-2-{7B, 13B, 70B}, GPT-J-6B, and Falcon-{7B, 40B}) varying in size and support for parallelism, quantization, and pruning (see supplementary material).    
Unless noted otherwise, we deploy an LLM every five minutes to simulate dynamic environments under varying loads.

\begin{figure*}[t]
    \centering
    \includegraphics[width=\linewidth]{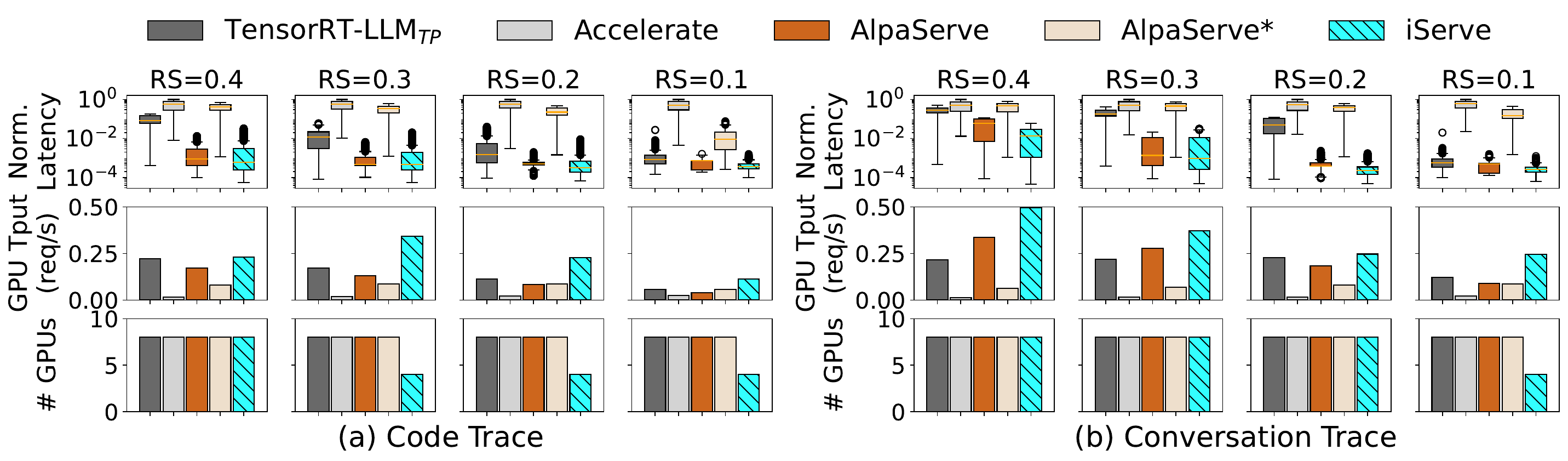}
    \vspace{-9mm}
    \caption{Given the intent to minimize latency, {\system} consistently minimizes latency (row 1), maximizes GPU throughput (row 2), and minimizes the \# of GPUs consumed (row 3) as the query arrival rate (rate scale factor) fluctuates under Azure's two LLM inference traces: (a) code and (b) conversation trace. See \S~\ref{sec:eval_e2e_lat}.
    }
    \label{fig:e2e_no_slo_lat_s1}
\end{figure*}

\noindent \textbf{Baselines.} We compare {\system} to four strong baselines: AlpaServe~\cite{li2023alpaserve}, Accelerate~\cite{sylvain2022accelerate}, and two baselines that use TensorRT-LLM~\cite{tensorrt-llm}, an advanced and widely-used LLM inference
serving framework. 

To simulate the behavior of non-expert users, 
we use two static baselines: (1) \emph{TensorRT-LLM$_{PP}$} which deploys LLMs with the maximum PP degree, and (2) \emph{TensorRT-LLM$_{TP}$} which deploys with the maximum TP degree.
Both do not employ quantization or pruning.
TensorRT-LLM$_{PP}$ follows the default configuration in the \texttt{transformers} library~\cite{huggingface}.

(3) \emph{Accelerate}~\cite{sylvain2022accelerate}, a library used by prominent LLM frameworks (e.g., HuggingFace~\cite{huggingface}), includes the Fully Shard\-ed Data Parallel (FSPD) algorithm~\cite{zhao2023pytorch}.
FSDP automates LLM deployment by sharding the LLM state across GPUs without applying quantization or pruning.

(4) \emph{AlpaServe}~\cite{li2023alpaserve}, a state-of-the-art LLM inference serving system, attempts to meet SLOs by auto-selecting parallelism for a set of LLMs to deploy on a fixed GPU cluster.
It requires a priori knowledge of request arrival patterns, an unrealistic assumption for general-purpose, dynamic inference-serving systems~\cite{Mark2019Zhang, Romero2021InFaaS}.
However, to ensure its best performance, we provide AlpaServe with our full trace beforehand and deploy its configurations (DP, TP, PP) on the same TensorRT-LLM platform as {\system}.


AlpaServe may skip deploying some LLMs to maximize system SLO attainment~\cite{mms_model_parallelism}.
Hence (5) \emph{AlpaServe*} is a slightly tweaked, more realistic version that does not drop LLMs.


\noindent\textbf{Testbed.} 
Our testbed resembles that of relevant prior work~\cite{agrawal2024tamingthroughputlatencytradeoffllm, fu2024serverlessllm, Miao2024SpecInfer}. 
We use 8 $\times$ NVIDIA RTX A6000 (48GB) GPUs. 
Pairs of GPUs are connected with NVLink and the remaining connections are PCIe.
The server has an AMD EPYC 7763 CPU (64 cores).
We use Ubuntu 22.04 and NVIDIA Driver 535.171.04 with CUDA Version 12.2.

\noindent\textbf{Metrics.}
We evaluate {\system} 
using (a) metrics capturing each user intent (e.g., latency, total memory), (b) GPU throughput, and (c) the number of GPUs used.
We also report SLO attainment---the percentage of SLOs we meet---when {\system} is given latency SLOs.

\noindent\textbf{Results.} 
Highlights of our results are:
\vspace{-2mm}
\begin{itemize}[leftmargin=*]
    \item Given the intent to minimize latency, {\system} reduces latency by 77.62\% while improving GPU throughput by 4.72$\times$ on average across loads compared to baselines (\S~\ref{sec:eval_e2e_lat}).
    \item Given the intent to minimize cost, memory, or GPU-hours, {\system} reduces cost by 86.70\%, memory consumption by 75.97\%, and GPU-hours by 86.31\% on average across loads compared to the baselines (\S~\ref{sec:eval_e2e_other_intents}).
  
    \item Given the intent to meet latency SLOs, {\system} improves SLO-attainment by 3.03-7.09$\times$ as the query arrival rate increases compared to AlpaServe (\S~\ref{sec:eval_e2e_slo}).
    \item {\system} best meets different user intents while reducing the profiling GPU hours by 6.05$\times$ on average compared to AlpaServe (\S~\ref{sec:eval_estimation_new}). 
    \item {\system} scales as the number of LLMs to deploy and cluster size grows, increasing the number of LLMs deployed by $2.58\times$ and cluster throughput by $1.30-11.11\times$ compared to baselines (\S~\ref{sec:eval_stress_test_new}).

\end{itemize}

\vspace{-4mm}
\subsection{iServe with the Minimize-Latency Intent}
\label{sec:eval_e2e_lat}

We evaluate {\system}'s efficacy in minimizing latency.
We deploy a selection of larger LLMs (1$\times$Llama-2-70B, 2$\times$Falcon-40B, and 1$\times$Llama-2-7B), however, we include additional experiments with other model sets.
Overall, {\system} consistently minimizes latency while maximizing GPU throughput and minimizing the number of GPUs used across varying loads, query patterns, and model sets.


\noindent\textbf{{\system} under consistent high load.}
We first evaluate {\system} under high load (rate scale factor 0.4) using the code trace (Figure~\ref{fig:e2e_no_slo_lat_s1}a column 1).
{\system} reduces median latency by 31.86\% compared to AlpaServe and by $>90\%$ compared to the other baselines.
Both TensorRT-LLM$_{TP}$ and Accelerate severely inflate latency due to poor configuration decisions: they both maximize parallelism for all LLMs.
TensorRT-LLM$_{TP}$ increases the small Llama-2-7B's latency by $1.54\times$ with TP=8 compared to {\system}'s TP=4, as the all-reduce overheads outweigh inference acceleration.
Accelerate's FSDP algorithm bloats latency due to network overheads from all-gather operations (e.g., Falcon-40B's latency is 94\% higher under Accelerate compared to {\system}).
Moreover, by maximizing parallelism, every GPU hosts all LLMs, increasing GPU congestion: queuing delay dominates inference latency for all LLMs under both systems.

AlpaServe autoselects LLM deployment configurations.
Both AlpaServe and {\system} deploy the LLMs with TP=4/PP=1, but AlpaServe's placement policy poorly packs them within four GPUs and replicates each LLM with DP=2 across the remaining GPUs.
This increases latency due to GPU contention.
Meanwhile, {\system} deploys the four LLMs across eight GPUs without applying DP with its hybrid, load-aware policy to reduce congestion.
Although AlpaServe's brute-force placement algorithm is load-aware (i.e., it is given the trace), it greedily skips over {\system}'s better placement decision in its search space navigation to reduce search time. 
Hence, AlpaServe increases median latency by 32\% compared to {\system}.
{\system}'s p99 latency is $2.37\times$ higher than AlpaServe, however, this is because AlpaServe drops Llama-2-70B to avoid congestion.
In a realistic setting where LLMs cannot be dropped, AlpaServe*'s poor placement decisions severely inflate p95 latency by 95.96\% compared to {\system}.


\noindent\textbf{{\system} under varying query arrival rates.}
Figure~\ref{fig:e2e_no_slo_lat_s1}a columns 1-4 shows that {\system} minimizes latency compared to the baselines regardless of the query arrival rate (load).
As the load decreases (rate scale 0.4 to 0.3), {\system}'s GPU throughput slightly grows (row 2) from 0.23 to 0.34.
With lower load, {\system}'s load-aware hybrid placement policy packs the LLMs onto 50\% fewer GPUs (four versus eight) without harming latency, increasing the amount of work each GPU completes.
TensorRT-LLM$_{TP}$ and Accelerate do not offer LLM-to-GPU placement policies; they simply use all eight GPUs regardless of the load.
AlpaServe's placement algorithm always consumes all GPUs by applying DP, however, this is unnecessary at low load.
{\system}'s per-LLM configuration decisions and load-aware placement policy enable it to intelligently pack LLMs onto fewer GPUs at lower loads, increasing GPU throughput by 1.03-18.49$\times$ while minimizing latency compared to the baselines.


\noindent\textbf{{\system} under varying query arrival patterns.}
We analyze {\system}'s robustness to varying query arrival patterns with Azure's conversation trace (Figure~\ref{fig:e2e_no_slo_lat_s1}b).
GPU throughput increases for all systems under this trace due to the consistent work provided with the trace's steady, high arrival pattern.
{\system} adapts its placement decisions to the arrival pattern to meet the user intent: at low load (0.2-0.3), {\system} disperses the LLMs across all GPUs (unlike the code trace) to prevent congestion, ensuring that {\system} consistently outperforms the baselines across query patterns and loads.

\begin{figure}[t]
    \centering
    \includegraphics[width=.99\linewidth]{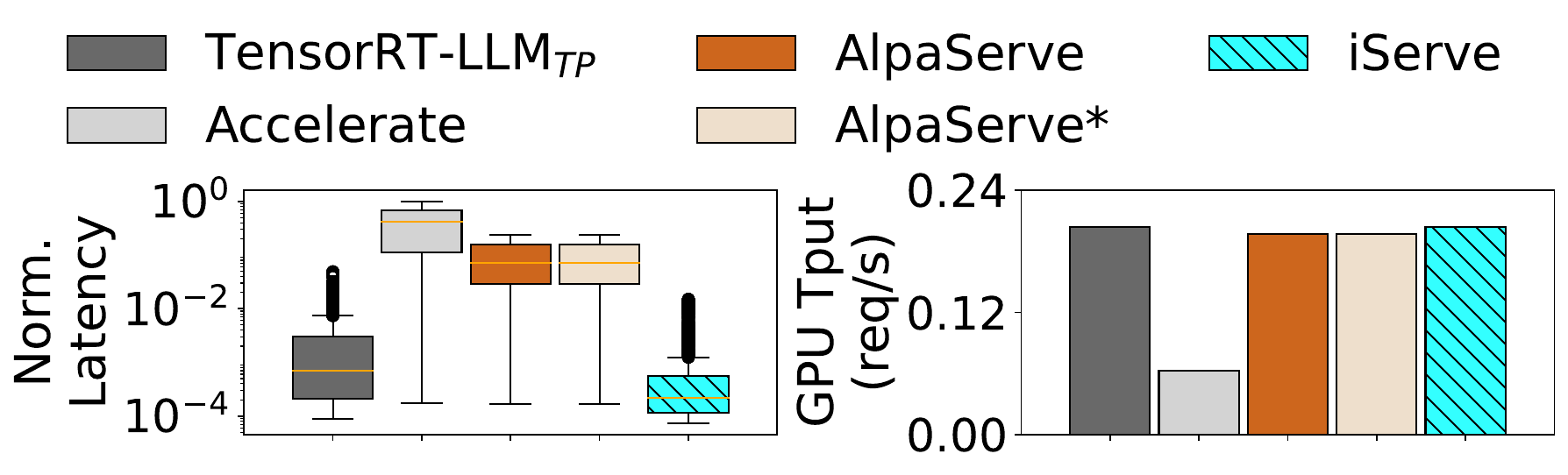}
    \vspace{-4mm}
    \caption{Latency and GPU throughput of {\system} when deploying several smaller LLMs. {\system} continues to minimize latency while maximizing throughput. See \S~\ref{sec:eval_e2e_lat}.
    }
    \label{fig:e2e_no_slo_lat_s2}
\end{figure}

\noindent\textbf{{\system} under varying lists of models.}
Figure~\ref{fig:e2e_no_slo_lat_s2} shows {\system}'s latency and GPU throughput as it deploys a larger set of smaller LLMs under high load with the code trace (findings hold for other loads and traces): 2$\times$Llama-2-13B, 2$\times$Llama-2-7B, 2$\times$Falcon-40B, and 2$\times$GPT-J-6B.
{\system} continues to minimize latency: it reduces p95 latency by 63.55-99.96\% compared to our baselines.
Unlike the previous set of LLMs, AlpaServe performs worse than TensorRT-LLM$_{TP}$ in latency.
By greedily skipping over potential configurations to reduce search time, AlpaServe chooses to deploy all LLMs with TP=2/PP=4, which is consistently worse than TensorRT-LLM$_{TP}$'s TP=8 configuration in minimizing latency (\S~\ref{sec:characterization} Figure~\ref{fig:tp-pp-curve}).
Interestingly, AlpaServe's GPU throughput is comparable to {\system} when deploying many small LLMs.
This is because (1) {\system}'s decisions ensure the LLMs serve queries faster than the rate of arrival (the request rate is not fast enough to demonstrate {\system}'s upper bound in throughput), and (2) the impact of AlpaServe's sub-optimal configurations is less pronounced with smaller LLMs, as the latency variation across a small LLM's configuration search space is small.
Nonetheless, {\system} is robust to any LLM type (small or large), as it makes accurate configuration decisions that meet the user's intent.

\begin{figure}[t]
    \centering
    \includegraphics[width=\linewidth]{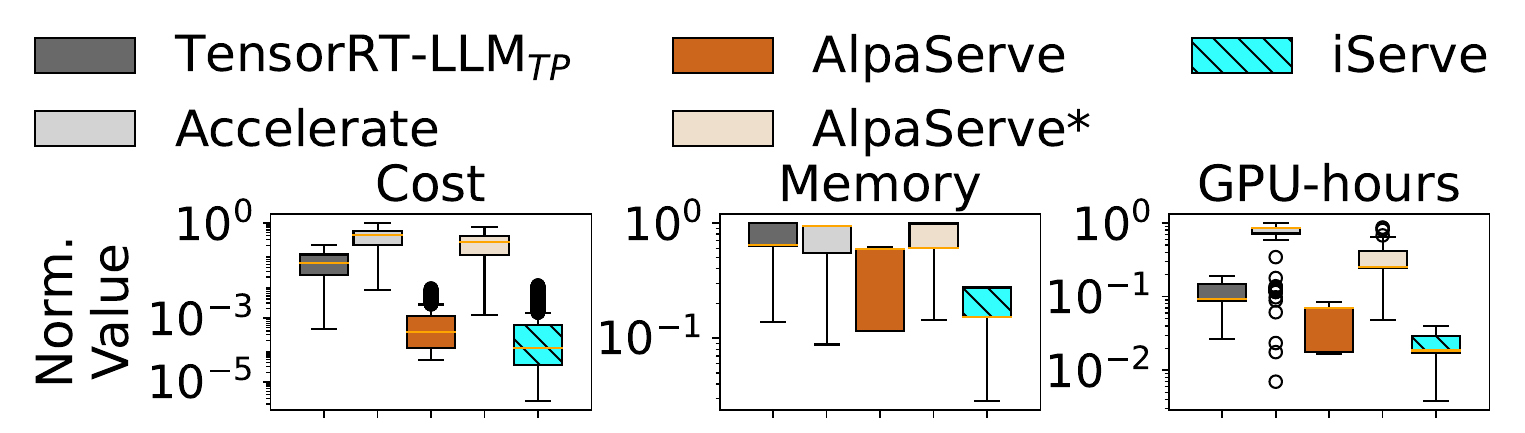}
    \vspace{-8mm}
    \caption{{\system} adapts to the intent, minimizing cost (col. 1), memory (col. 2), and GPU-hours (col. 3). See \S~\ref{sec:eval_e2e_other_intents}.
    }
    \vspace{-3mm}
    \label{fig:e2e_no_slo_other_s1}
\end{figure}

\vspace{-2mm}
\subsection{iServe with Other User Intents}
\label{sec:eval_e2e_other_intents}
\vspace{-1mm}
We next evaluate {\system}'s ability to meet diverse user intents (i.e., minimizing cost, memory, and GPU-hours).
Figure~\ref{fig:e2e_no_slo_other_s1} shows {\system}'s performance at high load under the code trace (see supplementary material A.2 for other traces and loads).
{\system} efficiently profiles and learns the impact of combinations of parallelism, quantization, and pruning on varying user intents.
For example, {\system} heavily quantizes (e.g., INT4, GPTQ4) and reduces parallelism to lower memory footprint by 74.36-83.92\% compared to the baselines.
Minimizing cost (memory $\times$ latency) and GPU-hours requires balancing resource allocation and latency.
However, TensorRT-LLM$_{TP}$ and Accelerate inflate both intents by increasing resource consumption (memory, GPUs), even for small LLMs that do not gain from additional resources.
AlpaServe buckets LLMs and makes configuration decisions per bucket, not \emph{per LLM}, rendering it inflexible to the intent.
In contrast, {\system} learns LLM-specific traits to adapt to the intent; for example, quantizing Llama-2 variants to INT4 reduces latency (and thereby cost and GPU-hours) by 59.14\% compared to FP16, as reduced-bit precision formats lower memory traffic and computational latency~\cite{dettmers2023case4bitprecisionkbit, wu2023understandingint4quantizationtransformer, yao2023zeroquantv2exploringposttrainingquantization}.

\begin{figure}[t]
    \centering
    \includegraphics[width=.9\linewidth]{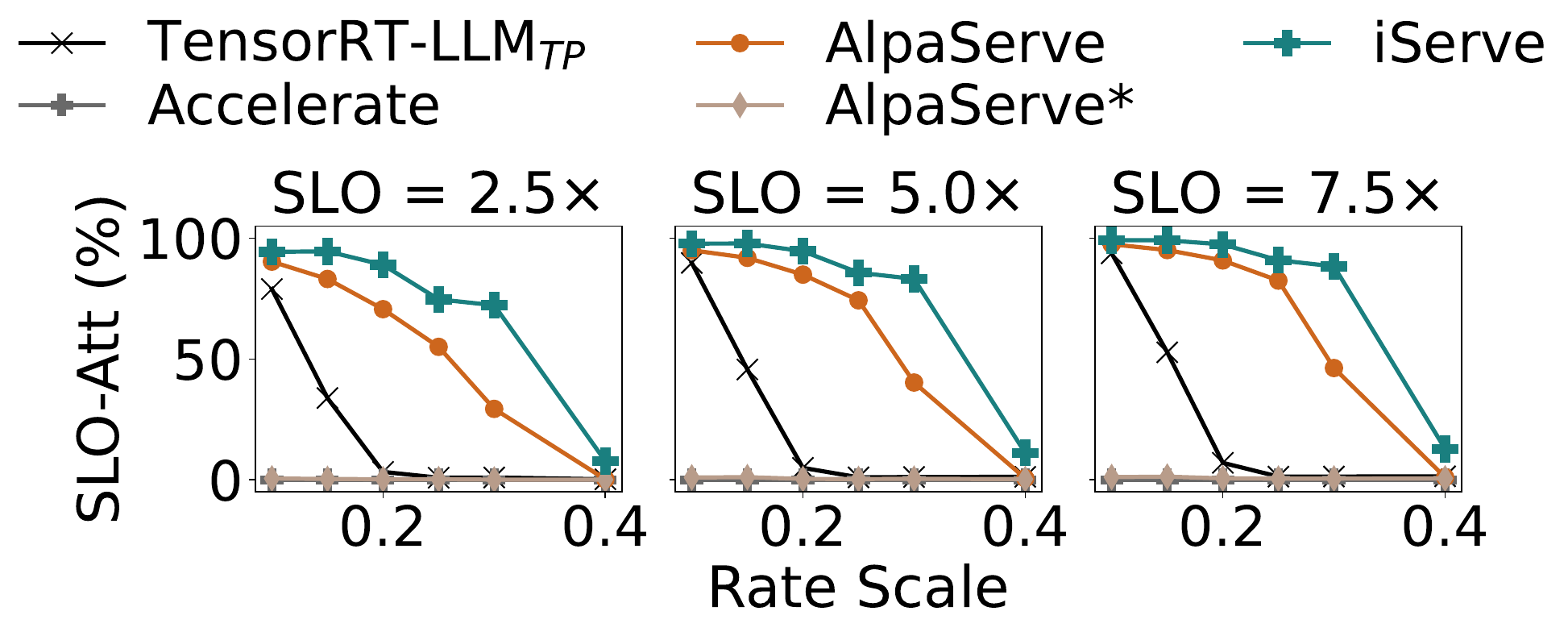}
    \vspace{-5mm}
    \caption{{\system} maximizes SLO attainment. See \S~\ref{sec:eval_e2e_slo}. 
    }
    \vspace{-3mm}
    \label{fig:e2e_slo}
\end{figure}

\vspace{-2mm}
\subsection{iServe Maximizes SLO Attainment}
\label{sec:eval_e2e_slo}
\vspace{-1mm}
We analyze {\system}'s efficacy in meeting latency SLOs.
Following Proteus~\cite{Ahmed2024Proteus}, we observe each LLM's latency under minimum parallelism without quantization and set SLOs as multiples of this latency (2.5-7.5$\times$).
Figure~\ref{fig:e2e_slo} shows {\system}'s SLO attainment across varying query arrival rates for a set of primarily large LLMs under both tight and loose SLOs. 

{\system} consistently outperforms the baselines in SLO attainment across different SLOs and query arrival rates.
TensorRT-LLM$_{TP}$ and Accelerate default to maximum parallelism, inflating SLO violations due to GPU contention and intermediate overheads (all-reduce, all-gather).
AlpaServe* consistently violates SLOs, as we restrict it from dropping LLMs during deployment.
Finally, despite a priori knowledge of the query arrival pattern, AlpaServe violates SLOs, as it (1) greedily skips configurations in its search space navigation to reduce search time, and (2) makes configuration decision per set of LLMs, not per LLM.

\begin{figure}[t]
     \centering
        \includegraphics[width=\linewidth]{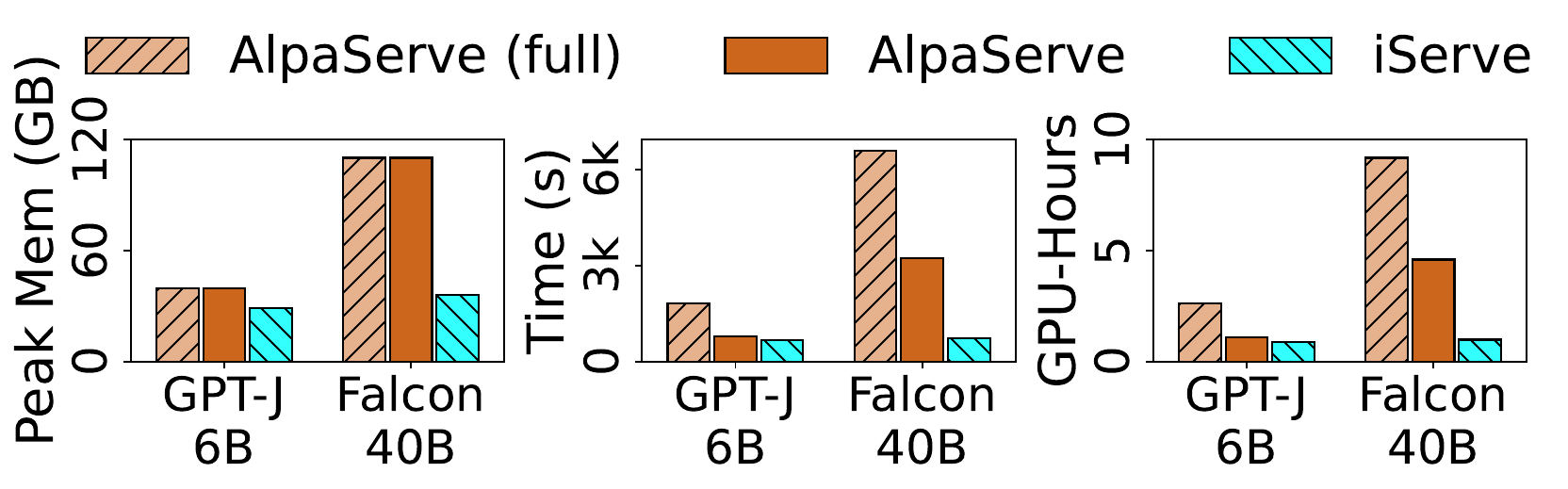}
        \vspace{-9mm}
        \caption{{\system} reduces LLM profiling cost (memory usage, time, and GPU-hours) compared to AlpaServe. See \S~\ref{sec:eval_estimation_new}. 
        }
        \vspace{-3mm}
        \label{fig:eva_prof_cost}
\end{figure}

\vspace{-3mm}
\subsection{{\system}'s Profiling Accuracy and Cost}
\label{sec:eval_estimation_new}

{\system}'s efficacy in meeting the user intent depends on 
{\system}'s fingerprint-based LLM profiling. 
We evaluate the accuracy and cost of {\system}'s profiling.



\noindent \textbf{Error analysis.} 
{\system} average latency and memory estimation error is $4.91\%$ ($<50$ms error) and $6.92\%$ (1.74GB), respectively (see A.2 in supplementary).
{\system}'s estimations are more precise for larger LLMs (e.g., Falcon-40B, Llama-2-70B), as errors for smaller models constitute a larger percentage of their overall footprint.
{\system}'s modular design enables cluster operators to easily use our more accurate profiling methods at the expense of profiling cost (\S~\ref{sec:latency-profiling_new}).


\noindent \textbf{Cost analysis.}
Figure \ref{fig:eva_prof_cost} presents AlpaServe's and {\system}'s profiling cost for a small (GPT-J-6B) and large (Falcon-40B) LLM (findings are representative of other LLMs).
As AlpaServe only navigates model parallelism, we also implement support for quantization and pruning with AlpaServe (full) to show its profiling cost under the entire search space.
AlpaServe's profiling memory, time, and GPU-hours are 1.37 – 3.04$\times$, 1.16 – 4.36$\times$, and 1.22 – 4.60$\times$ higher, respectively, compared to {\system}.
Adding support for quantization and pruning only increases its profiling cost: AlpaServe (full)'s profiling time and GPU-hours are 2.70–8.92$\times$ and 2.89–9.20$\times$ higher compared to {\system}.
Although {\system} navigates a larger search space, it cuts profiling cost by observing lightweight LLM fingerprints on a subset of configurations, all while meeting user intents and maximizing GPU throughput.

\begin{figure}[t]
    \centering
    \includegraphics[width=\linewidth]{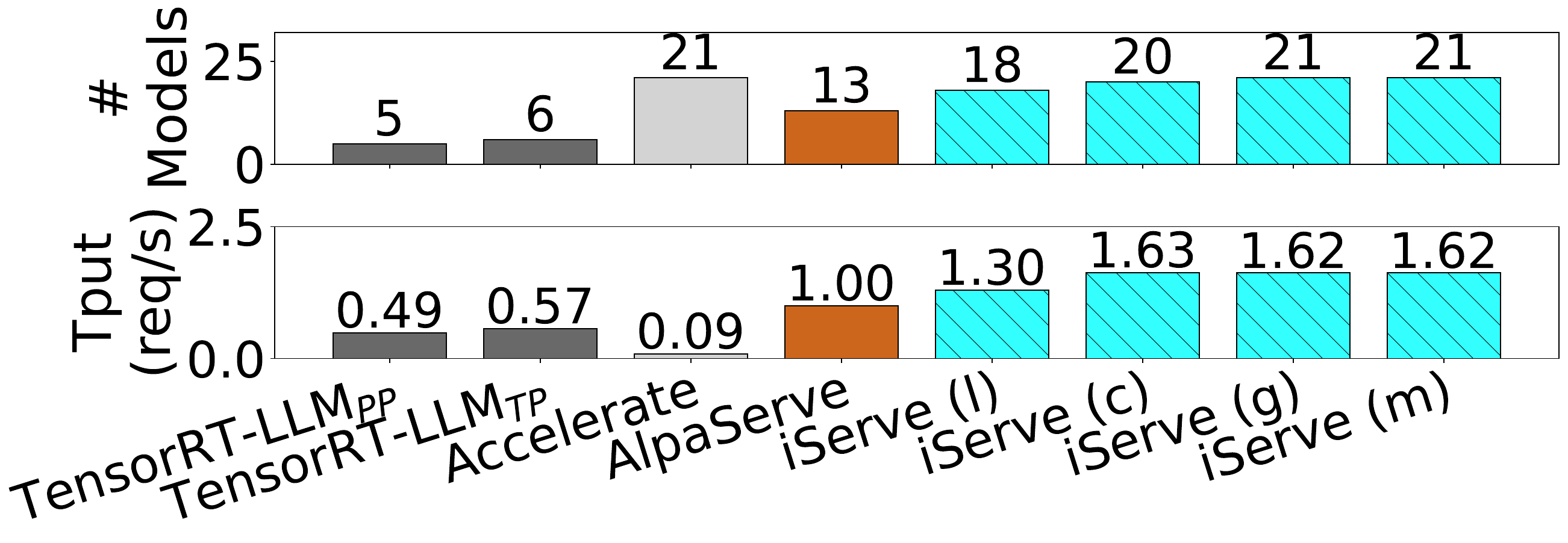}
    \vspace{-8mm}
    \caption{{\system} scales and maximizes the cluster throughput and number of LLMs deployed in a fixed cluster size. {\system} sustains high performance across intents: minimize latency (l), cost (c), GPU-hour (g), and memory (m). See \S~\ref{sec:eval_stress_test_new}.
    }
    \vspace{-2mm}
    \label{fig:e2e_max_load}
\end{figure}

\vspace{-3mm}
\subsection{{\system} Scales with LLMs and Cluster Size}
\label{sec:eval_stress_test_new}

\noindent\textbf{Scale with number of LLMs.} 
To demonstrate {\system} scales with the number of LLMs, we use a set of 21 LLMs.
We synthesize a one-hour trace by merging two Azure conversation traces with medium load.
We deploy two LLMs every three minutes and randomly assign queries to LLMs.

Regardless of the intent, {\system} scales effectively as the number of LLMs grows, boosting cluster throughput by 
$1.30$-$18.11\times$ compared to the baselines (Figure~\ref{fig:e2e_max_load} row 2).
{\system} deploys all 21 LLMs when minimizing memory or GPU-hours by aggressively quantizing (INT4) and minimizing parallelism to reduce overheads.
For latency and cost minimization, {\system} deploys 18 and 20 LLMs, respectively, as {\system} increases parallelism and quantizes less. 
{\system} scales by not only making accurate configuration decisions, but also by dynamically adapting its configurations and LLM layer partitioning across GPUs based on the available GPU memory.
For example, when minimizing latency, {\system} deploys the last Falcon-7B instance with PP=4 to lower its per-GPU memory requirement and distributes its layers unevenly across 4 GPUs (10, 9, 7, 6 layers) to fit within available resources.
Though this configuration is not optimal, {\system} chooses the best-effort configuration to adapt to remaining resources.
We note that Accelerate serves all 21 LLMs by offloading weights to CPU DRAM but suffers from PCIe transfer overheads, degrading GPU throughput.

\noindent\textbf{Scale with cluster size.}
(a) In a homogeneous cluster (e.g., all A6000 GPUs), {\system}'s profiling cost remains constant, as the configuration search space does not expand.
Although more GPUs present an opportunity to increase parallelism, such large parallelism (e.g., TP$>8$) is rarely used, as the all-reduce and all-gather overheads outweigh latency benefits~\cite{shoeybi2020megatronlmtrainingmultibillionparameter, li2023alpaserve}.
We observe diminishing returns with larger parallelism (\S~\ref{sec:characterization} Figure~\ref{fig:tp-pp-curve}): Falcon-40B's latency reduces by 39.75\% when TP increases from 2 to 4, but TP=8 only cuts it by 11.02\%.
(b) In a heterogeneous cluster (e.g., A6000s, A100s, H100s), {\system}'s profiling cost increases with each GPU type.
Though the search space remains constant, latency and memory profiles vary by GPU type, requiring profiling for each.
However, {\system}'s fingerprint profiling strategy keeps costs manageable.
To reduce cost, we could extrapolate profiles across GPU types by adjusting for architectural differences; we leave this for future work.



\vspace{-2mm}
\subsection{{\system}'s Profiling using Harvested Resources}
\label{sec:eval_overhead_new}
{\system}'s fingerprint-driven profiling is lightweight. Hence, we harvest the spare GPUs in the serving cluster when feasible. 
We evaluate the impact of such harvesting.
We deploy multiple LLMs on each GPU and dispatch a high query load to them.
We profile Llama-2-70B while LLMs serve requests from the conversation trace at varying loads and assess the impact of harvesting on SLO attainment.
We omit the figure due to space constraints (see A.2 in Supplementary).
{\system} profiles the large Llama-2-70B with minimal service disruption when load $<30\%$, reducing SLO attainment by just 4.00\%. 
Under higher loads ($>30\%)$, SLO attainment drops below 90\%.
Ultimately, this is not a limitation of {\system}, but rather the cluster size to load ratio.
It is at the discretion of cluster operators using {\system} to decide whether to harvest GPU resources or use dedicated GPUs for profiling based on load.
Even in the latter case, {\system}'s fingerprints reduce profiling time and resource requirements (\S~\ref{sec:eval_estimation_new}).

\vspace{-3mm}
\section{Discussion}
\label{sec:discussion}
\vspace{-1mm}

{\system}'s modular design enables extensibility to new user intents, models, and hardware.

\noindent \textbf{Extensibility to new intents.} 
{\system} can effortlessly support new intents derived from latency and memory data (e.g., throughput). 
To accommodate broader intents (e.g., energy efficiency, carbon reduction), the Controller can be extended to profile additional metrics (e.g., power consumption), though further work is needed to accurately extrapolate these from the fingerprint to the full LLM. 
These techniques can still be easily integrated into {\system}.

\noindent \textbf{Extensibility to new models.} 
{\system} seamlessly supports LLMs that follow the commonly used decoder-only architecture. 
To accommodate models that deviate from this structure (e.g., LLaVA~\cite{liu2023llava}, which includes image layers), the Fingerprint Controller can be extended to profile unique layers for those models to enable lightweight profiling.

\noindent \textbf{Extensibility to new hardware.}
{\system} supports homogeneous clusters regardless of the GPU type (e.g., A100s, A6000s), as its fingerprint and load-aware placement policy treat hardware as an opaque box.
But, deploying LLMs across heterogeneous devices introduces new challenges, such as latency variation across LLM stages deployed on different GPU types. We leave this investigation to future work.

\vspace{-3mm}
\section{Related Work}
\label{sec:related_work}

\noindent\textbf{LLM serving systems.} 
The rise of LLMs has led to the development of various LLM serving systems~\cite{holmes2024deepspeedfastgenhighthroughputtextgeneration, prabhu2024vattentiondynamicmemorymanagement, liu2023dejavucontextualsparsity, agrawal2024tamingthroughputlatencytradeoffllm, zheng2024learn, xue2024moeinfinityoffloadingefficientmoemodel, Sheng2023flexgen, chen2024efficient, zhong2024distserve, pan2024instinfer, xue2024powerinfer, fu2024serverlessllm, 280922, 10.1145/3437801.3441578}.
Most of these enhance LLM operations (e.g. stall-free continuous batching~\cite{holmes2024deepspeedfastgenhighthroughputtextgeneration, agrawal2024tamingthroughputlatencytradeoffllm}, 
attention offloading~\cite{chen2024efficient}) or runtime scheduling~\cite{280922, agrawal2024tamingthroughputlatencytradeoffllm}, both of which are complementary to {\system}.
A few works attempt to navigate the parallelism search space for LLM deployment~\cite{li2023alpaserve, miao2023spotserve, Agrawal2024, agrawal2024etalonholisticperformanceevaluation, tan2024redcoastlightweighttoolautomate}.
SpotServe~\cite{miao2023spotserve} adapts LLM parallelism to spot instance availability; {\system}'s profiler can run atop SpotServe.
Vidur~\cite{Agrawal2024} uses simulation to predict optimal parallelism and scheduling strategies.
AlpaServe~\cite{li2023alpaserve} uses a greedy ILP algorithm to make parallelism decisions; we show {\system}'s superior performance (\S~\ref{sec:evaluation_new}).
These systems require expensive profiling, code changes, and/or future query knowledge. {\system} makes (1) intent-aware deployments via efficient search with {\submodel}s and (2) load-aware LLM-to-GPU placement without foreseeing future load.

\noindent\textbf{LLM frameworks.} 
Several frameworks facilitate and optimize LLM training and inference~\cite{rajbhandari2020zeromemoryoptimizationstraining, aminabadi2022deepspeed, shoeybi2020megatronlmtrainingmultibillionparameter, nanotron, gale2022megablocksefficientsparsetraining, FairScale2021, paxml, mosaicml2022composer, openllm, llmfoundry, kwon2023vllm, tensorrt-llm, sylvain2022accelerate, rayllm, mlc-llm, saxml, yang2021mosec, article_flexmoe, zheng2023sparda}.
TensorRT-LLM~\cite{tensorrt-llm} is optimized for NVIDIA GPUs with custom attention kernels, in-flight batching, and paged KV-caching.
DeepSpeed~\cite{rajbhandari2020zeromemoryoptimizationstraining, aminabadi2022deepspeed} supports model parallelism and integrates high-performance inference kernels, communication optimizations, and heterogeneous memory technologies. 
vLLM~\cite{kwon2023vllm} leverages PagedAttention to optimize memory management.
{\system} build on top of these frameworks and can easily be ported.


\noindent\textbf{LLM inference optimization.} 
Several works propose optimizations to handle the scale of LLMs: model parallelism~\cite{shallue2019measuring, huang2019gpipe, Wang_2022, Singh_2023, zhao2023pytorch, 10098952, 9472938, Wagenlander2023TenplexDP, 10.1145/3620666.3651359} to distribute LLMs across GPUs, quantization~\cite{dettmers2023case4bitprecisionkbit, dettmers2022llmint8, frantar2023gptq, dettmers2023qlora, lin2023awq, int8kvcache, ma2024era, 2310.11453, Guo_2023}, pruning~\cite{ma2023llmpruner, sun2023wanda, frantar2023sparsegpt}, distillation~\cite{timiryasov2023baby, Hinton2015DistillingTK}, low-rank adaptation~\cite{hu2022lora}, or a combination of those~\cite{frantar2024marlin}.
These methods add to the configuration search space that {\system} efficiently navigates. 


\noindent\textbf{Resource allocation and multiplexing.} 
Several inference systems target traditional DNN models (e.g., Proteus~\cite{Ahmed2024Proteus}, InFaaS~\cite{Romero2021InFaaS}, Clipper~\cite{201468}, Clockwork~\cite{clockwork2020Gujarati}).
These systems dynamically allocate and auto-scale resources to improve latency and throughput.
However, their scaling and multiplexing techniques do not trivially extend to LLMs which demand substantial memory and compute.
Llumnix~\cite{sun2024llumnix} focuses on runtime scheduling, load balancing, and auto-scaling LLMs; these techniques are complementary to {\system}.


\vspace{-3mm}
\section{Conclusion}
\label{sec:conclusion}

We present {\system}, an intent-based LLM serving system that automatically deploys an LLM with a configuration 
that best meets the user intent.
{\system} efficiently navigates the large LLM deployment configuration space by (1) profiling an LLM's fingerprint under a few configurations and (2) using this data to estimate latency and memory footprint for all deployment options.
During deployment, {\system} uses estimated data to select the configuration that best meets user intent and is feasible to deploy given the GPU cluster's availability.
It then leverages its load-aware placement policy to deploy the LLM on specific GPUs.
Our experiments with various models, traces, and loads show that {\system} outperforms multiple state-of-the-art baselines in meeting user intent while increasing GPU throughput.

\bibliographystyle{plain}
\bibliography{reference} 

\clearpage  
\setcounter{page}{1} 
\appendix

\section{Supplementary Material}

\noindent This document comprises of supplementary material, from the Measurement Study (\ref{sec:app_measurement}) and Evaluation Experiments (\ref{sec:app_eval}) that were not shown in the paper due to space limitations.

\subsection{Supplementary Measurement Study}
\label{sec:app_measurement}

In this section, we provide the complete measurement study, which encompasses both quantization and pruning evaluations.

\begin{figure}[t]
    \centering
    \includegraphics[width=0.7\columnwidth]{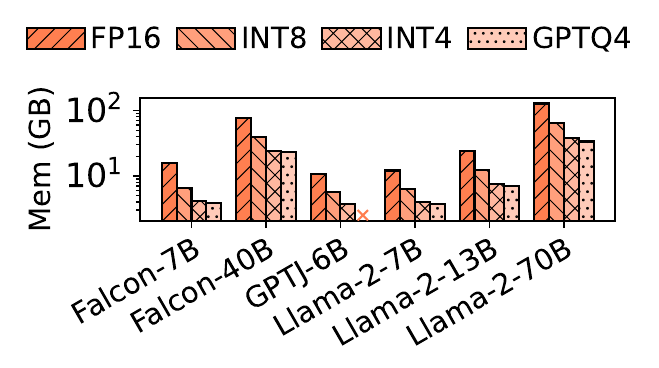}
    \vspace{-5mm}
    \caption{The effect of quantization on memory reduction becomes more significant as the level of quantization becomes more aggressive.}
    \label{fig:app_quantization}
\end{figure}

\noindent \textbf{Effect of quantization.} 
Figure~\ref{fig:app_quantization} presents the memory consumption for six LLMs across various quantization levels. 
The results indicate the memory footprint consistently decreases with more aggressive quantization.

\begin{figure}[t]
    \vspace{4mm}
    \centering
    \includegraphics[width=\columnwidth]{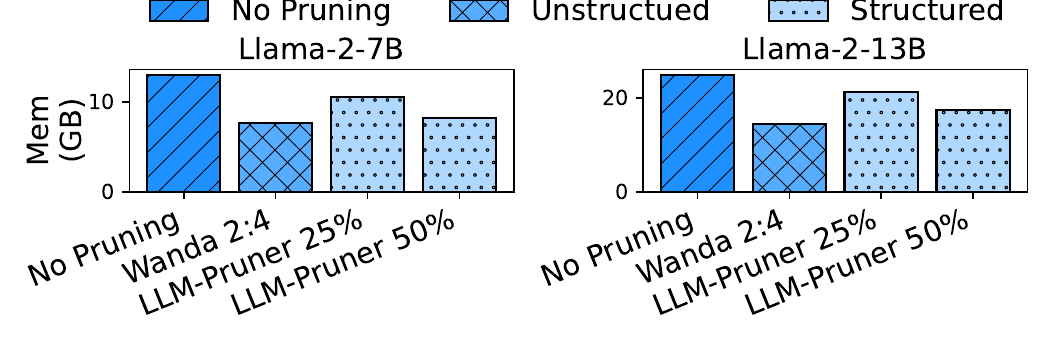}
    \vspace{-9mm}
    \caption{The memory of LLMs can be reduced with pruning techniques.}
    \label{fig:app_pruning}
\end{figure}

\noindent \textbf{Effect of pruning.} 
Figure~\ref{fig:app_pruning} illustrates the memory consumption for two LLMs subjected to different pruning techniques. 
The results demonstrate that pruning impacts memory, depending on the technique used.

\subsection{Supplementary Evaluation Experiments}
\label{sec:app_eval}
In this section, we present the complete evaluation results, which provide supporting and complementary evidence to the findings demonstrated for {\system} in the primary evaluation.

\noindent \textbf{List of LLMs.} Table \ref{tab:app_models} presents the six state-of-the-art LLMs used throughout the evaluation, which vary in size and support for parallelism, quantization, and pruning.

\begin{figure}[t]
    \vspace{-1mm}
    \centering
    \includegraphics[width=\linewidth]{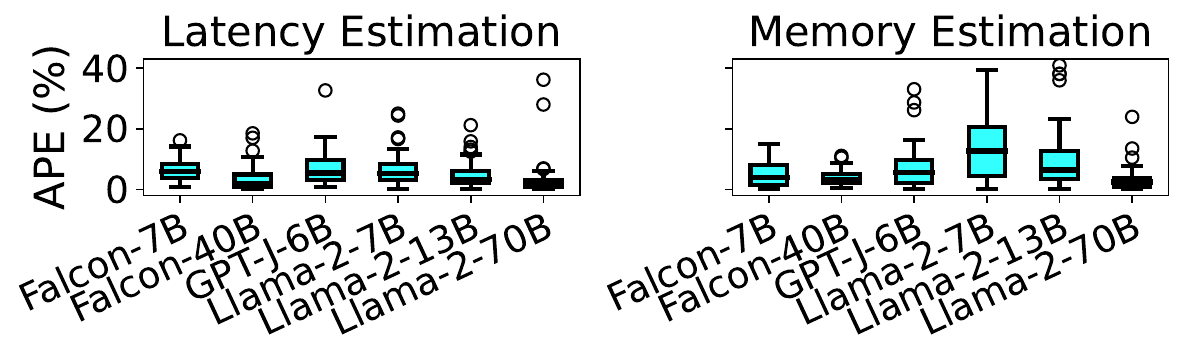}
    \vspace{-9mm}
    \caption{{\system} predicts the latency and memory requirements across deployment configurations with minimal error.}
    \label{fig:app_prof_error}
\end{figure}

\noindent \textbf{Profiling Accuracy.} 
Figure~\ref{fig:app_prof_error} shows that {\system} generally makes accurate latency and memory estimations per deployment configuration across LLMs.

\begin{figure}[t]
    \vspace{-3mm}
    \centering
    \includegraphics[width=0.8\linewidth]{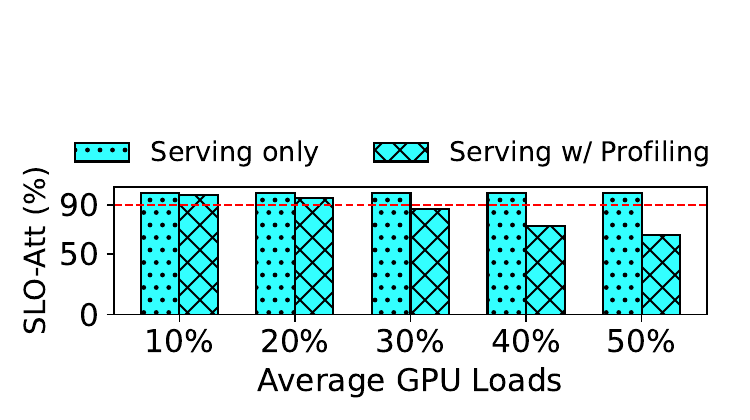}
    \vspace{-4mm}
    \caption{{\system} harvests resources for profiling while maintaining minimal impact on serving.}
    \label{fig:app_harvest}
\end{figure}

\noindent \textbf{Profiling using Harvested Resources.} 
Figure~\ref{fig:app_harvest} evaluates the profiling of a Llama-2-70B model while serving requests under various GPU loads. 
{\system} can maintain an SLO attainment of greater than 90\% when the average GPU load is below 30\%, but experiences higher violations under heavier loads.

\begin{figure*}[t]
    \centering
    \includegraphics[width=0.91\linewidth]{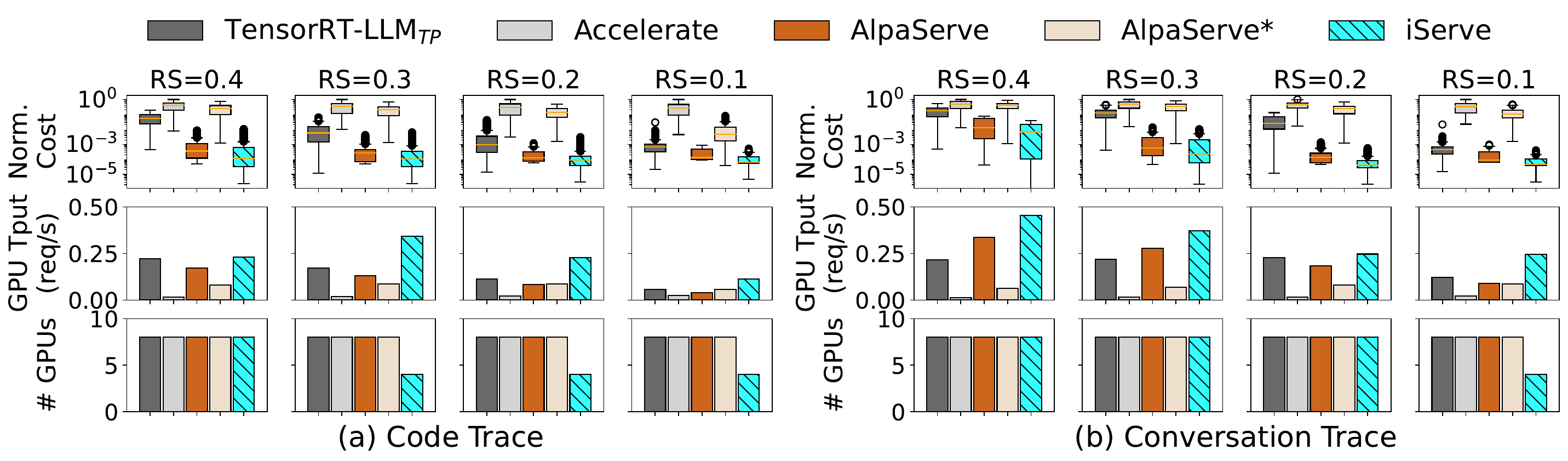}
    \vspace{-5mm}
    \caption{Given the intent to minimize cost, {\system} can minimize cost (row 1), maximize GPU throughput (row 2), and minimize \# GPU used (row 3) across all query arrival rates under both of Azure's traces: (a) code and (b) conversation.}
    \label{fig:app_E2E_NO_SLO_cost_S1}
\end{figure*}

\begin{figure*}[t]
    \centering
    \includegraphics[width=0.91\linewidth]{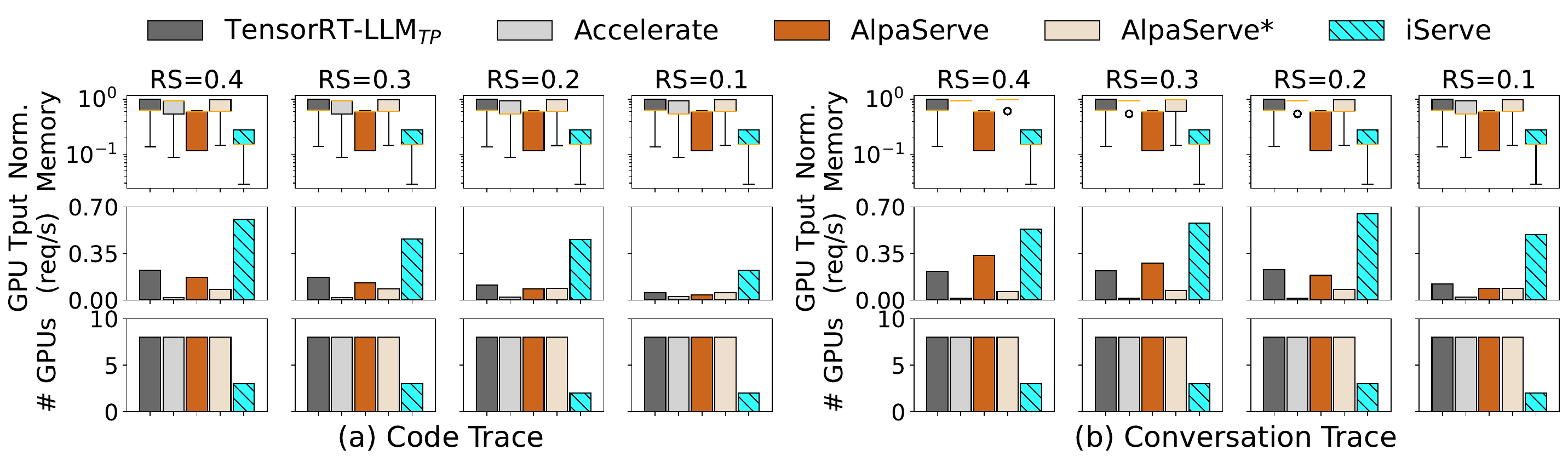}
    \vspace{-5mm}
    \caption{Given the intent to minimize memory, {\system} can minimize memory (row 1), maximize GPU throughput (row 2), and minimize \# GPU used (row 3) across all query arrival rates under both of Azure's traces: (a) code and (b) conversation.}
    \label{fig:app_E2E_NO_SLO_mem_S1}
\end{figure*}

\begin{figure*}[t]
    \centering
    \includegraphics[width=0.91\linewidth]{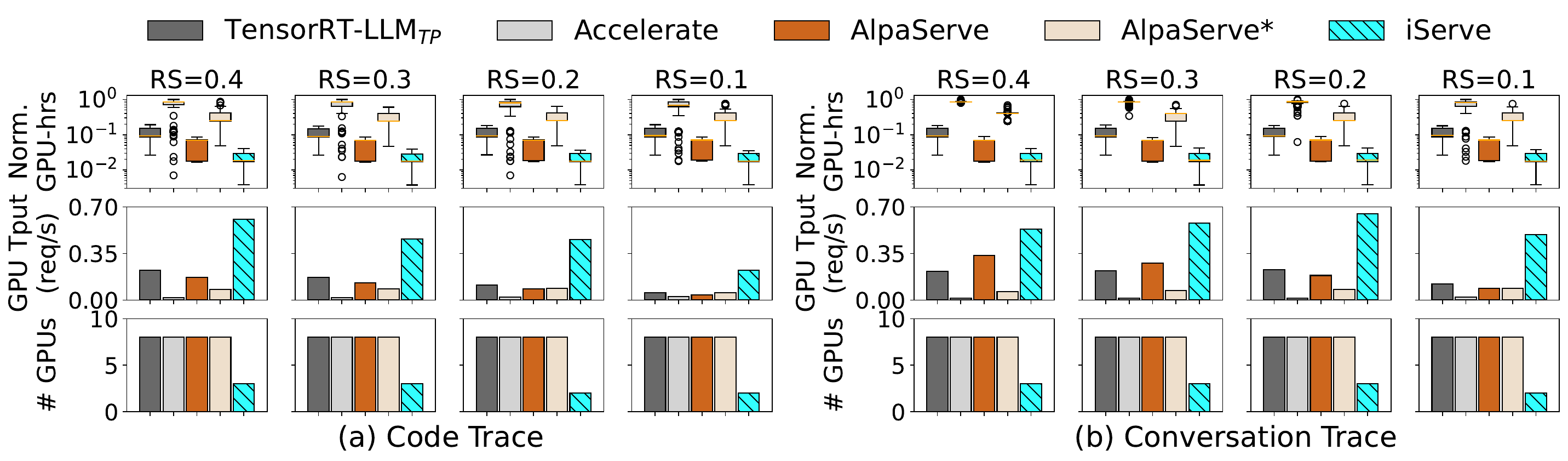}
    \vspace{-5mm}
    \caption{Given the intent to minimize GPU-hours, {\system} can minimize GPU-hours (row 1), maximize GPU throughput (row 2), and minimize \# GPU used (row 3) across all query arrival rates under both of Azure's traces: (a) code and (b) conversation.}
    \label{fig:app_E2E_NO_SLO_GPU-hours_S1}
\end{figure*}

\noindent \textbf{iServe with Other User Intents.} 
Figure~\ref{fig:app_E2E_NO_SLO_cost_S1},~\ref{fig:app_E2E_NO_SLO_mem_S1}, and~\ref{fig:app_E2E_NO_SLO_GPU-hours_S1} show the performance of {\system} serving large size LLMs (1$\times$Llama-2-70B, 2$\times$Falcon-40B, and 1$\times$Llama-2-7B) under various loads for the code trace and conversation trace, with the user intents of minimizing cost, minimizing memory, and minimizing GPU hours. 
{\system} makes advancements in meeting all user intents while maintaining high throughput and optimal GPU usage by leveraging parallelism, compression techniques, and load-balancing strategies.

\begin{figure*}[t]
    \centering
    \includegraphics[width=0.91\linewidth]{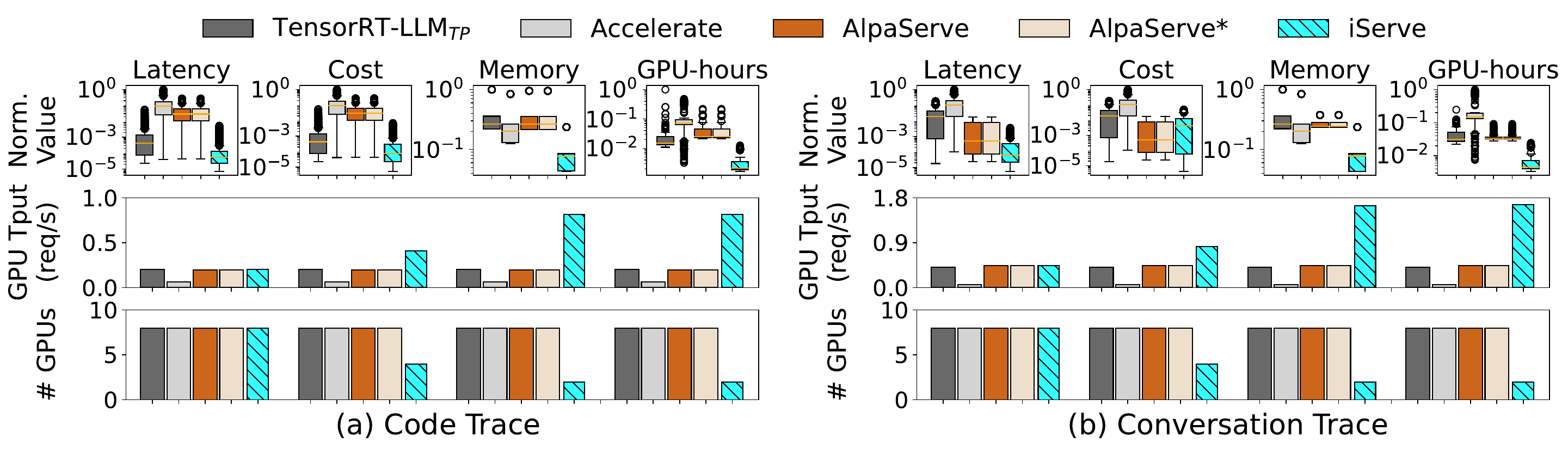}
    \vspace{-5mm}
    \caption{
    {\system} meets the intent to minimize latency, cost, memory, or GPU-hours in serving a large set of small LLMs.}
    \label{fig:app_E2E_NO_all_S2}
\end{figure*}

\noindent \textbf{iServe under varying lists of models.} 
Figure~\ref{fig:app_E2E_NO_all_S2} shows how {\system} meets all four user intents when serving a mixed set of large and small LLMs (2$\times$Llama-2-13B, 2$\times$Llama-2-7B, 2$\times$Falcon-40B, and 2$\times$GPT-J-6B). 
{\system} demonstrates robustness across various LLM types and is capable of making effective decisions when deploying a large number of models.

\newpage

{
\renewcommand{\arraystretch}{1.15} 
\begin{table}[t]
\vspace{4mm}
\centering 
\small
\begin{tabular}{|c|c|c|c|c|}
\hline
\begin{tabular}[c]{@{}c@{}} \textbf{Model} \end{tabular} &
\begin{tabular}[c]{@{}c@{}} \textbf{Parallelism}\end{tabular} &
\begin{tabular}[c]{@{}c@{}} \textbf{Quantization}\end{tabular} &
\begin{tabular}[c]{@{}c@{}} \textbf{Pruning} \end{tabular} \\
\hline
Falcon-7B    & Pipeline 
             & \multirow{3}{*}{INT8, INT4} 
             & \multirow{3}{*}{-} \\
\cline{1-1}\cline{2-2}
Falcon-40B   & \multirow{5}{*}{\begin{tabular}[c]{@{}c@{}} Pipeline,\\Tensor \end{tabular}}   &  &  \\ 
\cline{1-1}
GPT-J-6B     &  &  &  \\ 
\cline{1-1}\cline{3-3}\cline{4-4}
Llama-2-7B   &  & \multirow{3}{*}{\begin{tabular}[c]{@{}c@{}} INT8, INT4,\\GPTQ4,\\INT8 KV Cache \end{tabular}} 
                & \multirow{3}{*}{\renewcommand{\arraystretch}{0.8} 
                                  \begin{tabular}[c]{@{}c@{}} SparseGPT,\\Wanda,\\Wanda 2:4,\\Wanda 4:8 \end{tabular}} \\
\cline{1-1}
Llama-2-13B  &  &  &  \\
\cline{1-1}
Llama-2-70B  &  &  &  \\                    
\hline
\end{tabular}
\caption{Models and configurations used in the evaluation.}
\label{tab:app_models}
\end{table}
}

\end{document}